%% file: ms.tex
\documentclass[apj]{emulateapj}
\usepackage[normalem]{ulem}

\shorttitle{GCIRS~3 infrared spectra of H$_3^+$ and CO}
\shortauthors{Goto et al.}

\begin{document}

\title{Infrared H$_3^+$ and CO Studies of the Galactic Core:\\ GCIRS~3 and
  GCIRS~1W \altaffilmark{1}}

\author{Miwa Goto\altaffilmark{2},
        T. R. Geballe\altaffilmark{3},
        Nick Indriolo\altaffilmark{4},
        Farhad Yusef-Zadeh\altaffilmark{5},
        Tomonori Usuda\altaffilmark{6},
        Thomas Henning\altaffilmark{7},
and Takeshi Oka\altaffilmark{8}}

\email{mgoto@usm.lmu.de}

\altaffiltext{1}{Based on data collected during the CRIRES
  Science Verification program (60.7A-9057) and open time program
  (079.C-0874) at the {\it VLT} on Cerro Paranal (Chile), which
  is operated by the European Southern Observatory (ESO). Based
  also on data collected at Subaru Telescope, which is operated
  by the National Astronomical Observatory of Japan.  Based on
  data obtained from the ESO Science Archive Facility under
  request number MGOTO 43308.}

\altaffiltext{2}{Universit\"ats-Sternwarte M\"unchen,
Scheinerstr. 1, D-81679 Munich, Germany; mgoto@usm.lmu.de.} 

\altaffiltext{3}{Gemini Observatory,
                 670 North A`ohoku Place, Hilo, HI 96720; tgeballe@gemini.edu}

\altaffiltext{4}{Department of Physics and Astronomy, Johns
  Hopkins University, 3400 N. Charles St., Baltimore, MD 21218;
indriolo@pha.jhu.edu.}

\altaffiltext{5}{Department of Physics and Astronomy, Northwestern
University, Evanston, IL 60208; zadeh@northwestern.edu.}

\altaffiltext{6}{Subaru Telescope, 650 North A`ohoku Place, Hilo,
                  HI 96720; usuda@naoj.org.}

\altaffiltext{7}{Max-Planck-Institut f\"ur Astronomie,
  K\"onigstuhl 17, D-69117 Heidelberg, Germany;  henning@mpia.de.}

\altaffiltext{8}{Department of Astronomy and Astrophysics,
                 Department of Chemistry, and Enrico Fermi Institute,
                 University of Chicago, Chicago, IL 60637; t-oka@uchicago.edu}

\begin{abstract}

We have obtained improved spectra of key fundamental band lines of 
H$_3^+$, $R$(1,1)$^l$, $R$(3,3)$^l$, and $R$(2,2)$^l$, and 
ro-vibrational transitions of CO on sightlines toward the luminous 
infrared sources GCIRS~3 and GCIRS~1W, each located in the Central 
Cluster of the Galactic center within several arcseconds of Sgr A*.  
The spectra reveal absorption occurring in three kinds of gaseous 
environments: (1) cold dense and diffuse gas associated with foreground 
spiral/lateral arms; (2) warm and diffuse gas absorbing over a wide and 
mostly negative velocity range, which appears to fill a significant 
fraction of the Galaxy's Central Molecular Zone (CMZ); and (3) warm, 
dense and compact clouds with velocities near $+$50~km~s$^{-1}$ probably 
within 1-2 pc of the center. The absorptions by the first two cloud 
types are nearly identical for all the sources in the Central Cluster, 
and are similar to those previously observed on sightlines from Sgr A* 
to 30 pc east of it.  Cloud type (3), which has only been observed 
toward the Central Cluster, shows distinct differences between the 
sightlines to GCIRS~3 and GCIRS~1W, which are separated on the sky by 
only 0.33~pc in projection.  We identify this material as part of 
an inward extension of the Circumnuclear Disk previously known from HCN 
mapping. Lower limits on the products of the hydrogen ionization rate 
$\zeta$ and the path length $L$ are 2.3~$\times$~10$^5$~cm~s$^{-1}$ and 
1.5~$\times$~10$^3$~cm~s$^{-1}$ for the warm and diffuse CMZ gas and for 
the warm and dense clouds in the core, respectively. The limits indicate 
that the ionization rates in these regions are well above 
10$^{-15}$~s$^{-1}$.

\end{abstract}

\keywords{Galaxy: center --- ISM: clouds --- ISM: lines and bands ---
  ISM: molecules --- stars: individual (GCIRS~1W, GCIRS~3)}

\section{Introduction}

Detailed knowledge of the gaseous environment in the Galactic
center (GC) is of vital importance in understanding the wide
range of extraordinary phenomena that have recently taken place
and are currently taking place there. For several decades many
well-known atomic and molecular species have been used as probes
to help characterize that environment. In the last decade, a
newly discovered species, H$_3^+$, found in interstellar clouds
eighteen  years ago \citep{geb96} and
towards the Galactic center one year later \citep{geb99},
has proven to be a unique and valuable  tool in this
endeavor.

\begin{figure*}
\plotone{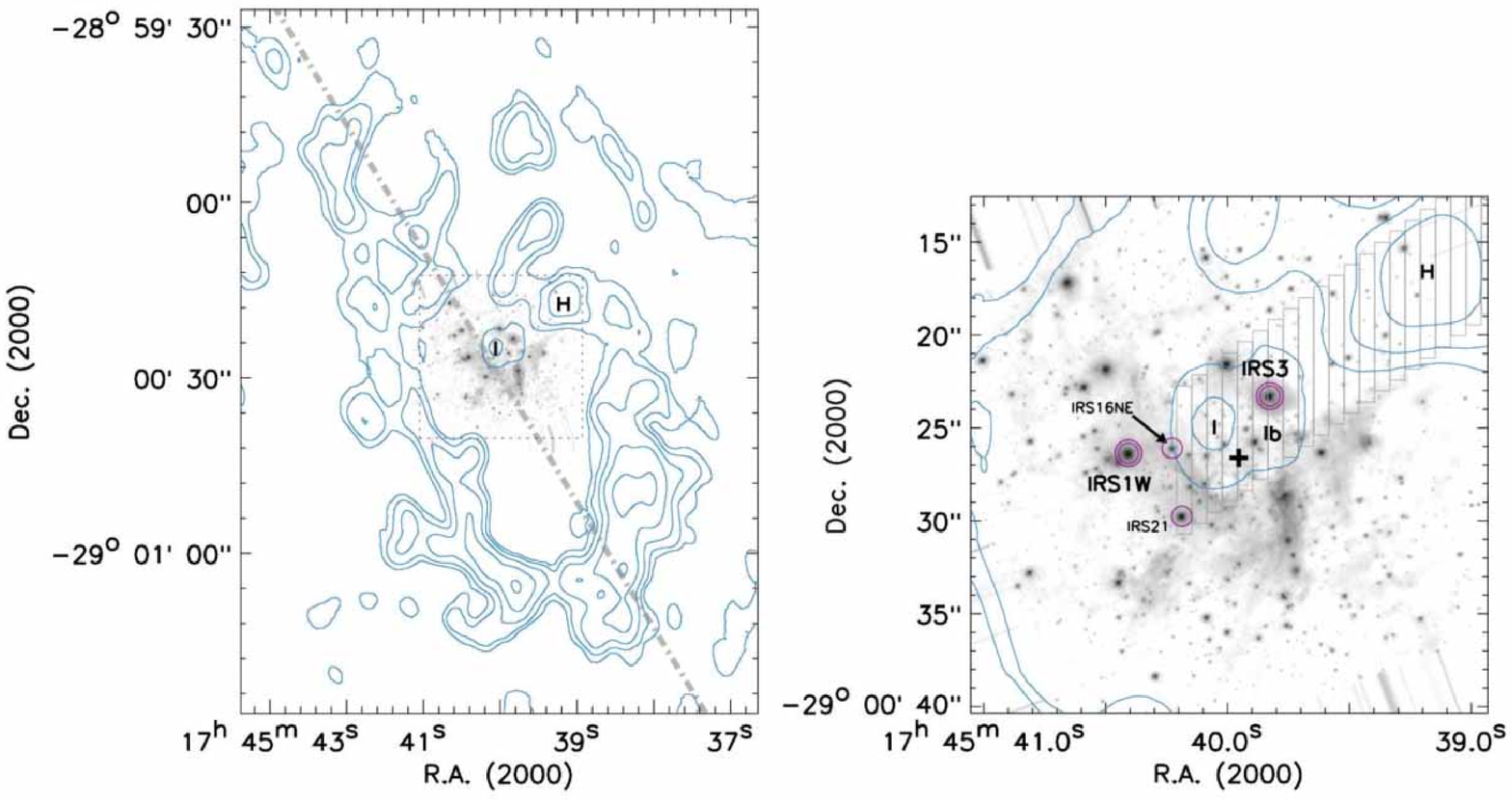}
\caption{$K$-band image of the Central Cluster obtained from ESO
  NACO archive data with contours of HCN $J$=4-3 emission 
  \citep{mon09} from the CND overlaid in blue. Locations of
  clumps I and H of \citet{mon09} discussed in Section
  4.3.2. are labeled. In the left panel the dot-dashed line
  is parallel to the Galactic plane and the area
  shown in the right panel is denoted by a dotted square
  \citep[originally published in][]{got13}. The locations of the
  infrared sources are marked with circles in the right panel
  and extraction apertures of the HCN spectra in
  Fig.~\ref{hcn_line} are shown as gray rectangles.
\label{hcn_image}}
\end{figure*}

The simple chemistry of H$_3^+$, i.e., its production through ionization 
of H$_2$ (mainly either by cosmic rays or X-rays) followed by the rapid 
proton hop reaction H$_2$~+~H$_2^+$~$\rightarrow$~H$_3^+$~+~H and its 
destruction, in diffuse clouds by recombination with electrons 
H$_3^+$~+~e~$\rightarrow$~H~+~H~+~H or H~+~H$_2$ and in dense clouds by 
reaction primarily with CO, allows one to reliably determine the product 
$\zeta$$L$, where $L$ is line of sight dimension of the cloud, and 
$\zeta$ is the ionization rate of H$_2$. In addition, column densities 
of H$_3^+$ in its low rotational levels are useful for measuring the 
temperature $T$ and number density $n$ of the gas 
\citep{oka04,oka05,oka06}.

Previous studies of H$_3^+$ in the GC have revealed the presence
there of a vast amount of warm ($T$ $\sim$ 250~K) and diffuse
($n$ $\leq$100~cm$^{-3}$) gas \citep{oka05,got08} and an
unusually high ionization rate, on the order of
$\zeta~\sim~10^{-15}$ s$^{-1}$. The presence of so much H$_3^+$
requires a drastic change in our understanding of the physical
states and volume filling factors of the gas in the Central
Molecular Zone \citep[CMZ, ][]{mor96,laz98}, the name given to
the region of radius $\sim$200 pc at the center. The high
ionization rate required to account for the large column density
of H$_3^+$ is potentially important for understanding the
overall energetics of the Galactic center.
A high ionization fraction and an elevated gas temperature
\citep[e.g.,][]{mor83}, much higher than the dust temperature of
$\sim$20~K \citep{pie00,lis01}, also may be consistent with the
cosmic rays being the dominant heating source for the gas
\citep{gus81,yus07b}.  Whatever its cause, the high gas kinetic
temperature may help to explain the current low star-forming
efficiency in the CMZ \citep{yus07b,an11,imm12} and the top-heavy
initial mass function observed in the Central Cluster
\citep{bar10} due to the associated reduction of ambipolar
diffusion and increase of the Jeans mass.

Of the previously studied 15 lines of sight within 3~pc of the
Galactic plane and longitudinally distributed from the Central
Cluster, within 1~pc of the central black hole, Sgr~A*, to the
Quintuplet Cluster 30~pc east of Sgr A* ($l\sim$0.18\degr)
\citep[][Oka et al.  unpublished]{got08}, the one toward
GCIRS~3, located only a few arcseconds from Sgr~A*
(Fig.~\ref{hcn_image}), is exceptional. While all 15 sightlines
show prominent absorption by H$_3^+$ in its $R$(1,1)$^l$ and
$R$(3,3)$^l$ lines, indicating high H$_3^+$ populations both in
the lowest ($J,K$) = (1,1) level and the (3,3) metastable level
361~K above ground, the $R$(2,2)$^l$ line is prominent only on
the sightline toward GCIRS~3, over a velocity range centered
near $+$50~km~s$^{-1}$ \citep{got08}. Since the (2,2) level
spontaneously decays to ground (1,1) \citep{pan86} with a
lifetime of 27~days \citep{nea96}, the density in the gas
producing this absorption must be considerably higher than in
other regions of the CMZ observed to date. This gas is likely to
be in the form of a compact cloud, since the sightline toward
GCIRS~1W, which is only 8\farcs4 away from GCIRS~3
\citep[0.33~pc, assuming equal radial distances of
  8~kpc;][]{eis03,ghe08}, produces a much weaker $R$(2,2)$^l$
absorption line.

In this paper we present, analyze, and discuss new and improved
spectra of H$_3^+$ toward GCIRS~3 and GCIRS~1W, as well as CO
ro-vibrational transition spectra toward them and few other
objects in the Central Cluster and the Quintuplet Cluster. The
velocity resolution of the CO spectra is higher by factors of
$\sim$5  to  $\sim$100 than previous studies of
the CO fundamental band lines toward these objects
  \citep{geb89,mon01,mou09}. The ratio of the column densities
$N$(CO)/$N$(H$_3^+$) is invariably much lower in diffuse clouds
($n$ $<$ 300~cm$^{-3}$) than dense clouds. Thus, observations of
both species help to discriminate between dense and diffuse
environments.

\section{Observations}

The observations consisted of high resolution infrared
spectroscopy of several Galactic center sources using
spectrographs at the Very Large Telescope (VLT) at Paranal in
Chile and the Subaru Telescope on Mauna Kea in Hawai`i. A
summary of the observations is given in Table~\ref{t1}.

\input{t1.tex}  

\subsection{CRIRES/VLT}

Spectra of H$_3^+$ lines toward GCIRS~3 and GCIRS~1W were obtained 
using CRIRES \citep{kau04} at the VLT on several occasions in June and 
August 2007 in an open time program (079.C-0874). The CRIRES 0\farcs2 
wide slit, oriented at position angle 113\degr\ so that GCIRS~3 and 
GCIRS~1W could be observed simultaneously, provided a velocity 
resolution of 3~km~s$^{-1}$. The adaptive optics system MACAO 
\citep{bon04} was used with a $R=13.5$~mag star 17\arcsec~distant from 
GCIRS~1W as the wavefront reference. The $R$(1,1)$^l$ 
(3.71548~$\mu$m), $R$(2,2)$^l$ (3.62047~$\mu$m), and $R$(3,3)$^l$ 
(3.53366~$\mu$m) lines were observed separately using three grating 
settings.

Spectra of CO toward GCIRS~3, GCIRS~1W, GCIRS16~NE, GCIRS~21
(all members of the Central Cluster; Fig.~\ref{hcn_image}), and
the Quintuplet Cluster source GCS~3-2 were obtained between
October 2006 and September 2007 in service observing mode during
science verification time for CRIRES (60.A-9057) as well as in
the open time program (079.C-0874), using the same adaptive
optics set-up as described above and an $R=15$~mag star
8\arcsec~away from GCS~3-2 as a wavefront reference for that
object. The $^{12}$CO $v$=2-0 overtone band, which has been used
by us previously \citep{oka05} was observed for GCIRS16~NE,
GCIRS~21, and GCS~3-2, but not for GCIRS~3 and GCIRS~1W, since
those objects are faint in the $K$-band; for them the
fundamental band $v$=1-0 was observed.  For the overtone band, the
spectral range 2.292--2.356~$\mu$m was covered in two grating
settings, allowing measurements of all $R$-branch lines and the
$P$(1)--$P$(4) lines. For the fundamental band, the spectral
range 4.692--4.809~$\mu$m was observed in two grating settings
covering the $P$(3) to $P$(15) lines of $^{12}$CO $v$=1-0 and
the $R$(9)--$R$(0) and $P$(1)--$P$(4) lines of $^{13}$CO
$v$=1-0. The $^{12}$CO $v$=1-0 lines are heavily saturated and
have not been analyzed, unless otherwise noted. The 0\farcs4
slit was used for GCIRS~3 and GCIRS~1W while the 0\farcs2 slit
was used for the other objects, resulting in velocity
resolutions of 6~km~s$^{-1}$ and 3~km~s$^{-1}$,
respectively. The early-type star HR~6879 (B\,9.5III,
$R$=1.85~mag) was observed before or after each GC object as a
telluric standard.

Spectra of GCS~3-2 were also obtained by CRIRES in the intervals 
2.200--2.255~$\mu$m and 2.087--2.133~$\mu$m to search for absorption 
from two quadruple transitions of H$_2$, $v$=1-0 $S$(0) and $v$=1-0 
$S$(1), in the same open time program. The 0\farcs2 slit was employed, 
resulting in a velocity resolution of 3~km~s$^{-1}$.

Spectra were extracted from the CRIRES data using the
crires\_spec\_jitter recipe\footnote{CRIRES Pipeline User Manual
  VLT-MAN-ESO-19500-4406.} on the ESO gasgano
platform\footnote{http://www.eso.org/sci/data-processing/software/gasgano/.}. The
results were consistent with the pre-processed CRIRES spectra
provided by the observatory. In some cases the latter spectra
were cosmetically better and were used for
analysis. Custom-written IDL code was employed to divide spectra
of the GC sources by the spectra of the telluric standard to
remove atmospheric absorption lines. Wavelength calibration was
obtained by cross-correlating the telluric absorption lines with
model atmospheric transmission spectra computed using ATRAN
\citep{lor92}. The uncertainty in the wavelength calibration
depends on the density of telluric lines and is typically less
than one pixel ($\leq$ 1~km~s$^{-1}$). The IRAF\footnote{IRAF is
  distributed by the National Optical Astronomy Observatories,
  which are operated by the Association of Universities for
  Research in Astronomy, Inc., under cooperative agreement with
  the National Science Foundation.}  {\it rv} package was used
to convert the observed wavelengths to velocity with respect to
the local standard of the rest utilizing the IAU definition of
the Sun's peculiar motion.

\subsection{IRCS/Subaru}

The spectrograph IRCS at the Subaru Telescope was used to obtain a 
spectrum of GCS~3-2 in the vicinity of the H$_2$~$v$=1-0~$S$(0) line 
on 24 May 2003 UT. The slit width was 0\farcs15, resulting in a 
velocity resolution of 15~km~s$^{-1}$. The adaptive optics system was 
used employing the same wavefront reference as for the CRIRES observations 
of GCS~3-2. The slit was oriented east-west. HR~7557 ($R$=0.77~mag, 
A7\,V) and HR~7121 ($R$=2.02~mag, B2.5\,V) were observed as telluric
standard stars. Reduction of the IRCS data was performed in the 
similar manner as for the CRIRES data except that the IRAF aperture 
extraction package was used to extract the spectra.

\section{Results}

\subsection{H$_3^+$}

\begin{figure}
\plotone{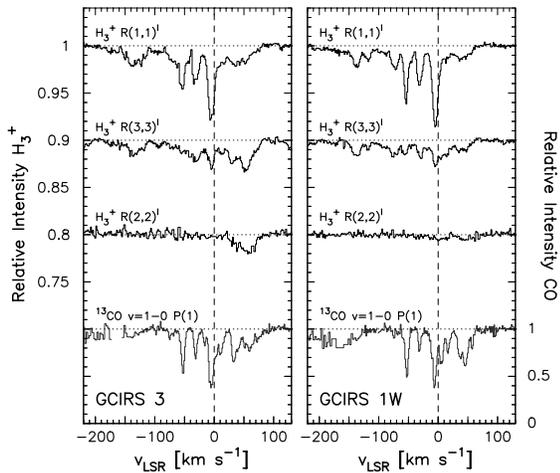}
\caption{Velocity profiles of H$_3^+$ $R$(1,1)$^l$,
  $R$(3,3)$^l$, $R$(2,2)$^l$, and $^{13}$CO $v$=1-0 $P$(1) lines
  toward GCIRS~3 (left) and GCIRS~1W (right).
\label{f1}}
\end{figure}

The spectra of the three H$_3^+$ lines toward GCIRS~3 and GCIRS~1W are 
shown in Fig.~\ref{f1} together with the spectrum of the $v$=1-0 $P$(1) 
line of $^{13}$CO. The $P$(1) line was selected for comparison because 
it is least affected by nearby telluric absorption lines and 
interstellar $^{12}$CO lines, as discussed in Section 3.2. The systemic 
radial velocity of GCIRS~1W is $+$35$\pm$20~km~s$^{-1}$ \citep{pau06}, 
measured by \ion{He}{1} absorption line at 2.06~$\mu$m \citep{pau04}. 
The radial velocity of GCIRS~3 is not known due to the absence of 
photospheric lines in its spectrum.

\subsubsection{Negative velocities}

At negative velocities the spectra in Fig.~\ref{f1} are similar
to those on the sightlines toward GCS~3-2 \citep[][located in
  the Quintuplet Cluster]{oka05} and NHS~21, NHS~22, NHS~25, and
NHS~42 \citep[][located between the Quintuplet and the Central
  Clusters]{got08}. Our interpretation of them is similar to those 
authors. The three sharp absorption components at $-$53~km~s$^{-1}$, 
$-$32~km~s$^{-1}$, and 0~km~s$^{-1}$, conspicuous in both the H$_3^+$ 
$R$(1,1)$^l$ and $^{13}$CO $P$(1) profiles, are due to relatively cold 
and dense gas in the lateral arms at 3~kpc and 4~kpc approaching the 
sun, and to foreground spiral arms, respectively. The same sets of the 
absorption lines have been observed in CO~$J$=3-2 
\citep{sut90,mon01} and NH$_3$ \citep{ser86} toward the Galactic 
center. In common with the previously observed sightlines from the 
Central Cluster to 30~pc east, absorption by CO 2--0 band lines at
negative velocities is almost entirely due to gas in these arms. In 
contrast, at negative velocities the H$_3^+$ $R$(1,1)$^l$ spectrum 
shows a broad and structured absorption trough upon which the three 
sharp absorptions due to the arms are superimposed. No absorption in the 
$R$(2,2)$^l$ line is present, indicating that the density of the gas 
producing the absorption troughs in the $R$(1,1)$^l$ and $R$(3,3)$^l$ line 
profiles is low. We therefore identify the absorption trough with 
diffuse gas in the CMZ, as in \citet{oka05}.

\begin{figure*}
\plotone{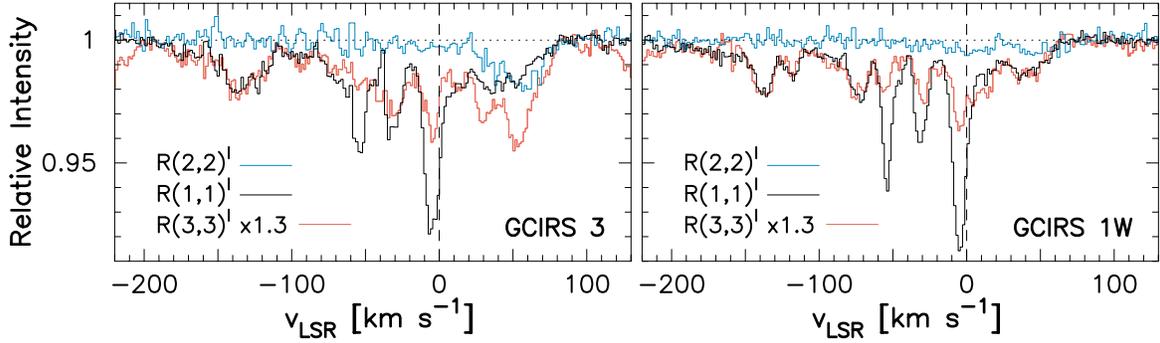}
\caption{Left: Absorption velocity profiles of H$_3^+$
  $R$(1,1)$^l$ (black), $R$(3,3)$^l$ (red), and $R$(2,2)$^l$
  (blue) toward GCIRS~3. $R$(3,3)$^l$ absorption is scaled by
  the factor of 1.3 to match the broad absorption trough of the
  $R$(1,1)$^l$ line. Right: Same, but for GCIRS~1W.
\label{f2} }
\end{figure*}

It is noteworthy that at negative velocities the $R$(3,3)$^l$ 
absorption profile, which contains no spiral arm components, matches 
the shape of the $R$(1,1)$^l$ trough. This similarity is clearly seen 
in Fig.~\ref{f2} in which the spectra of the $R$(3,3)$^l$ line, 
multiplied by a factor of 1.3 for each object, are superimposed on the 
spectra of the $R$(1,1)$^l$ line. This likeness has been noted earlier 
\citep{oka05,got08}, but has not previously been seen in such detail 
as in Fig.~\ref{f2}. It provides conclusive evidence that the trough 
in the $R$(1,1)$^l$ spectrum and the $R$(3,3)$^l$ spectrum arise in 
the same warm and diffuse gas. Indeed the accurate match indicates 
remarkably that although the gas that produces the absorption trough 
covers a wide range of velocities and presumably thus exists over a 
wide range of line of sight locations within the CMZ, its temperature 
is nearly constant. 

At negative velocities the $R$(3,3)$^l$ profiles, and in particular the 
one toward GCIRS~1W, also contain several weak and narrow absorption 
features. Some of these are close to the velocities of the foreground 
arms and might be interpreted as indicating the presence of warm gas in 
the spiral arms. We do not adopt this interpretation in view of the 
absence of such features in the $R$(3,3)$^l$ spectra on other sightlines 
and because close examination reveals that the velocities of the 
absorption peaks are not exact matches to those of the spiral arms (as 
observed in the CO $P$(1) and H$_3^+$ $R$(1,1)$^l$ line profiles).  It 
may be that they arise in compact and warm regions of the Circumnuclear 
Disk \citep[CND;][]{gen87}, the chain of molecular clouds orbiting 
Sgr~A* at $v$~$\sim$~100~km~s$^{-1}$ at a radius of 1.5~pc 
\citep{zha09}.
 
\subsubsection{Positive velocities}

At positive velocities the absorption profiles of the H$_3^+$ lines 
toward GCIRS~3 and GCIRS~1W are qualitatively different in two ways from 
those toward the other GC sources outside the Central Cluster shown in 
\citet{oka05} and \citet{got08}. First, absorption toward other GC stars 
is only present to about +30~km~s$^{-1}$, whereas absorption extends to 
$+$60~km~s$^{-1}$ toward GCIRS~1W and to $+$80~km~s$^{-1}$ toward 
GCIRS~3, in both the $R$(1,1)$^l$ and the $R$(3,3)$^l$ lines 
(Fig.~\ref{f1}). Note that these velocity extents are also present in 
the $^{13}$CO line in this figure. Second, the sightline toward GCIRS~3 
also produces strong and broad absorption at positive velocities in the 
$R$(2,2)$^l$ line (from $+$20 to $+$70~km~s$^{-1}$, peaking at 
$+$52~km~s$^{-1}$). The only other sightline for which this line has 
been seen strongly is that toward 2MASS J17470898-2829561, which is 
apparently located within the Sgr~B molecular cloud \citep{geb10,got11}, 
far removed from the central few parsecs. The $R$(3,3)$^l$ profile 
toward GCIRS~3 has two absorption maxima, at $+$28~km~s$^{-1}$ and 
$+$52~km~s$^{-1}$, whereas the $R$(2,2)$^l$ profile has only a single 
maximum near $+$52~km~s$^{-1}$.

Only a marginal detection of the $R$(2,2)$^l$ absorption near 
$+$40~km~s$^{-1}$ is evident toward GCIRS~1W (Fig.~\ref{f1}, right). 
Assuming that both GCIRS~1W and GCIRS~3 are members of the Central 
Cluster (see discussion below), the large differences in the H$_3^+$ and 
$^{13}$CO line profiles at positive velocities, on sightlines separated 
by 0.33~pc, suggests that the cloud (or clouds) producing the 
$R$(2,2)$^l$ absorption toward GCIRS~3 is compact, with linear dimension 
comparable to the sightline separation, and is located close to 
the Central Cluster.

\subsubsection{Equivalent widths and column densities}

As noted previously, apart from velocities near those of the spiral 
arms ($-$62$\rightarrow$ $-$45~km~s$^{-1}$, $-$37 $\rightarrow$ 
$-$25~km~s$^{-1}$, and $-$15 $\rightarrow$ $+$12~km~s$^{-1}$) and at 
positive velocities for GCIRS~3, a nearly constant scaling factor of 
1.3 exists between the $R$(3,3)$^l$ and $R$(1,1)$^l$ absorption 
profiles, as shown in Fig.~\ref{f2}. To estimate the strengths of the 
$R$(1,1)$^l$ absorptions originating in the CMZ in the above velocity 
intervals, and thus to estimate the total column densities of H$_3^+$ 
in the CMZ, we assume the same factor also applies in the above 
velocity intervals, multiply the $R$(3,3)$^l$ line profiles by 1.3, 
and use them as surrogates for the $R$(1,1)$^l$ CMZ absorptions toward 
these objects. This appears to be a more accurate method than the 
one used by \citet{oka05}, which did not include the velocity 
structure in the trough that is now evident in the $R$(3,3)$^l$ 
absorption profile observed by CRIRES.

\input{t2.tex} 

The equivalent widths of those portions of the H$_3^+$ lines
arising in the CMZ, together with the corresponding column
densities of the lower levels of the transitions, are listed in
Table~\ref{tb2} over several sub-intervals covering the entire
CMZ velocity range. The column densities were calculated using
$N$(H$_3^+$)$_{level} = $($3hc/8 \pi^3\lambda$)$W_\lambda /
|{\bf \mu}|^2$, where $|{\bf \mu}|^2$, the square of the dipole
moment, is 0.0141~D$^2$, 0.0177~D$^2$, and 0.0191~D$^2$ for the 
$R$(1,1)$^l$, $R$(2,2)$^l$, and $R$(3,3)$^l$ transitions, 
respectively. The Einstein A coefficients, given in \citet{nea96}, 
were also used in our previous studies \citep{got02,oka05,got08}. 
Uncertainties in the equivalent widths were estimated from the 
standard deviations in the nearby continuum multiplied by the 
wavelength interval. To determine the H$_3^+$ absorption equivalent 
width and associated column density due to gas in the spiral arms 
toward each source, we subtracted the scaled $R$(3,3)$^l$ surrogates 
from the $R$(1,1)$^l$ absorption profiles. The resulting values are 
given in Table~\ref{tb3}.

\subsection{CO} \label{3.2}

\begin{figure}
\plotone{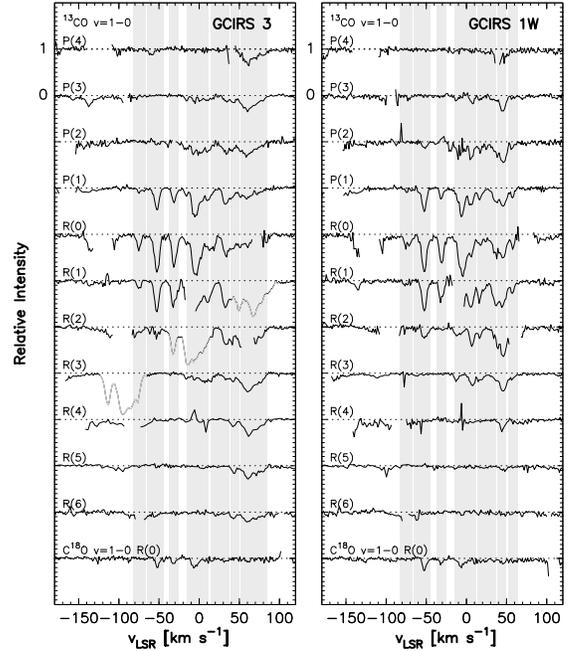}
\caption{Velocity profiles of $^{13}$CO~$v$=1-0
  $P$(4)--$P$(1),$R$(0)--$R$(6) and $^{18}$CO~$v$=1-0 $R$(0)
  absorption lines toward GCIRS~3 (left) and GCIRS~1W
  (right). Portions of the profiles contaminated by
  $^{12}$CO~$v$=1-0 are shown as gray lines. Velocity intervals
  used in the analysis are denoted by shaded vertical strips.
\label{co1}}
\end{figure}
\begin{figure}
\plotone{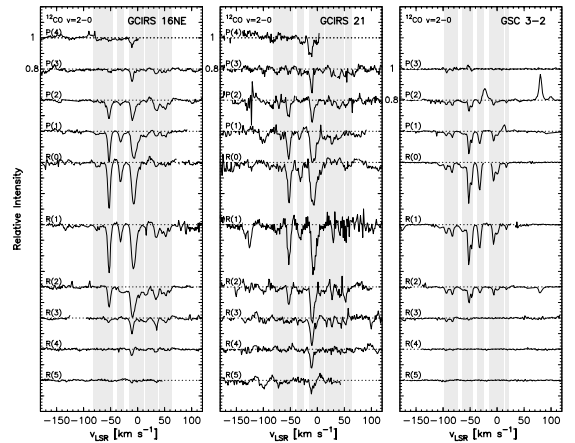}
\caption{Spectra of $^{12}$CO~$v$=2-0 $P$(3)--$P$(1) and
  $R$(0)--$R$(5) lines on sightlines to GCIRS~16NE, GCIRS~21,
  and GCS~3-2. Individual velocity components are shaded. 
  \label{co2}}
\end{figure}

\begin{figure}
 \plotone{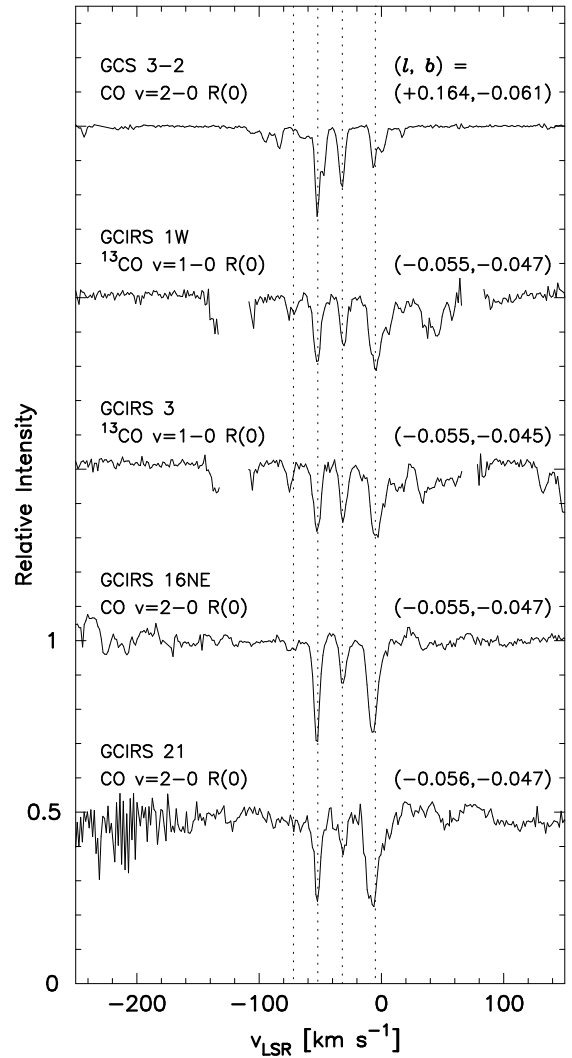}
 \caption{Velocity profiles of CO $R$(0) lines for five Galactic
   center sources (Galactic coordinates indicated on the
   right). Dashed lines from left to right correspond to
   radial velocities of an unidentified contributor at
   $-$72~km~s$^{-1}$, the 3 kpc expanding arm, the 4 kpc
   expanding arm, and clouds in more local arms as well as other
   clouds in circular orbits.
\label{r0}}
\end{figure}

Velocity profiles of lines of the $^{13}$CO $v$=1-0 fundamental band 
toward GCIRS~3, GCIRS~1W and of lines of the $^{12}$CO $v$=2-0 overtone 
band toward GCIRS~16NE, GCIRS~21, GCS~3-2 are shown in Figs.~\ref{co1} 
and \ref{co2}. In Fig.~\ref{r0} for each of these sources a profile of 
the $R$(0) line from either the 1-0 band of $^{13}$CO or from the 2-0 
band of $^{12}$CO is shown. In Figs.~\ref{co1} and \ref{co2} there are 
numerous gaps in the profiles due to strong telluric absorption lines. 
In addition some of the $^{13}$CO line profiles overlap with those of 
strong lines of the $^{12}$CO fundamental.  The $^{13}$CO $v$=1-0 
$R$($J$) lines nearly coincide with the C$^{18}$O $v$=1-0 $R$($J+$1) 
lines, with velocity offsets of 50--95~km~s$^{-1}$. The $^{13}$CO 
$v$=1-0 $P$($J$) lines nearly coincide with C$^{18}$O $v$=1-0 $P$($J-$1) 
lines with offsets of 18--30~km~s$^{-1}$. However in each case the 
contamination is relatively minor. C$^{18}$O v=1-0 $R$(0) is the
  only observed transition that is not contaminated by lines of
  $^{12}$CO and $^{13}$CO; a simple comparison of the line
  depths of C$^{18}$O $R$(0) and $^{13}$CO $R$(0) (the former
  shown at the bottom of Fig.~\ref{co1}) indicates that
  C$^{18}$O abundance is less than 1/4 of that of $^{13}$CO. The
  ratio above should be taken as an upper limit, since $^{13}$CO
  $R$(0) is expected to be more saturated than C$^{18}$O
  $R$(0).  

\input{t3.tex} 

\begin{figure*}
\plotone{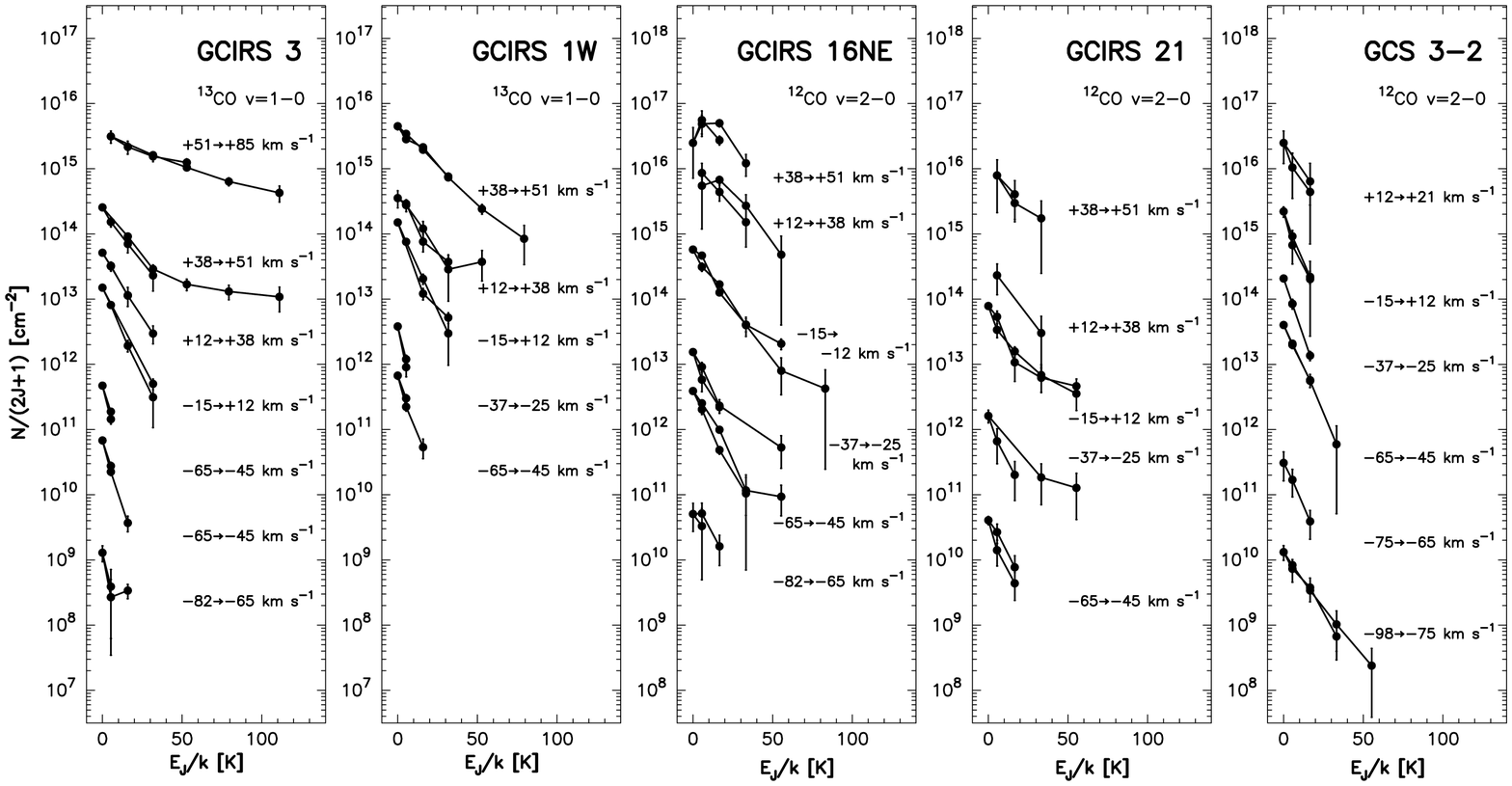}
\caption{CO population diagrams for four Central Cluster sources
  (GCIRS~3, GCIRS~1W, GCIRS~16NE, and GCIRS~21), and the
  Quintuplet Cluster source GCS~3-2, calculated from
  the spectra in Fig.~\ref{co1} and Fig.~\ref{co2} for the
  velocity intervals highlighted in those
  figures. Adjacent traces are offset vertically from one
  another by either 1 or 2 orders of magnitudes.
\label{pp}}
\end{figure*}

The measured equivalent widths and calculated column densities of the 
least contaminated $^{13}$CO $v$=1-0 and $^{12}$CO $v$=2-0 lines are 
listed in Tables~\ref{tb4} and \ref{tb5}. For the H$_3^+$ and $^{12}$CO 
$v$=2-0 lines, the absorption strengths $\Delta{I}$ and optical depths 
$\tau$ are approximately proportional, because the maximum optical 
depths are 0.1--0.3. For the fundamental band of $^{13}$CO, however, the 
non-linear equations $\Delta{I}(\lambda)=I_0(1-e^{-\tau(\lambda)})$ and 
$W_\lambda=\int\tau(\lambda)d\lambda$ must be used to relate the two, 
because the peak absorptions are as large as 80\%. The column densities 
of CO were calculated using the same equation as for H$_3^+$, but with 
$|{\bf
  \mu}|^2=\frac{S}{2J+1}(\mu_{1-0/2-0})^2$, where the spectral strength 
$S$ is $J+1$ for $R(J)$ lines and $J$ for $P(J)$ lines and 
$\mu_{1-0}$=0.1055~D, $\mu_{2-0}$=6.53$\times 10^{-3}$~D are the dipole 
moments of $^{12}$CO $v$=1-0 and $v$=2-0 transitions, respectively 
\citep{zou02}. The small differences between the transition dipole 
moments of $^{12}$CO~$v$=1-0 and $^{13}$CO~$v$=1-0 lines \citep{cha83} 
were neglected. Population diagrams, from which temperatures can be 
estimated, are shown in Fig.~\ref{pp} in velocity intervals 
corresponding to the various cloud and gas environments.

\input{t4.tex} 
\input{t5.tex} 


At negative velocities the dominant CO absorbers on all sightlines to 
the Central Cluster are the spiral and lateral arms.  An additional and 
weaker absorption at $-$72 km~s$^{-1}$ is present toward GCIRS~1, 
GCIRS~3, and GCIRS~16NE. In all of these the highest rotational level 
observed at negative velocities is $J$=3, and excitation temperatures 
range from 7 to 19~K (Table~\ref{t6}, and also Fig.~\ref{pp}), depending 
on the line of sight and the line pair. It is likely that the population 
distribution is subthermal \citep{neu12}.
The total column densities of $^{13}$CO in the three
  foreground arms were determined by summing the level
  populations using the values in Table~5 and are about
  $8~\times~10^{16}$~cm$^{-2}$) toward each source. This is in
  reasonable agreement with the value of $1.1 \times
  10^{17}$~cm$^{-2}$ of \citet{mon01} based on lower resolution
  ($R$=2000) spectroscopy of $^{13}$CO v=1-0 over an extended
  region near Sgr~A*.

\input{t6.tex} 

At positive velocities, as in the case of H$_3^+$ it seems
likely that most or all of the absorption by CO arises within
the CMZ. At 0~$<$~$v$~$<$~$+$38~km~s$^{-1}$ absorption toward
GCIRS~1W, GCIRS~3, and GCIRS~16NE is observed only up to
$J$=3. At higher positive velocities, a weak and narrow
$+$45~km~s$^{-1}$ absorption is observed toward GCIRS~1W out to
$J$=5, and a similar $+$43~km~s$^{-1}$ absorption is observed
toward GCIRS~3 out to $J$=6. Most strikingly, a strong and broad
absorption centered near $+$60~km~s$^{-1}$ is present toward
GCIRS~3 also up to $J$=6 for $^{13}$CO, and up to $J=$15
  for $^{12}$CO, but is not observed toward GCIRS~1W. The
velocity range of this feature is similar to those of the
positive velocity features seen toward GCIRS~3 in H$_3^+$ lines,
indicating that the absorptions by each species probably
originate in physically related gas components. The much higher
velocity gas ($\pm $300~km\,s$^{-1}$ or more) seen by
\citet{goi13} in the far-infrared emission lines of [\ion{N}{3}]
and [\ion{O}{1}] was not observed in any of the $^{13}$CO v=1-0
lines, nor even in the wings of the saturated $^{12}$CO v=1-0
lines.

\subsection{H$_2$}\label{3.3}

\begin{figure}
\plotone{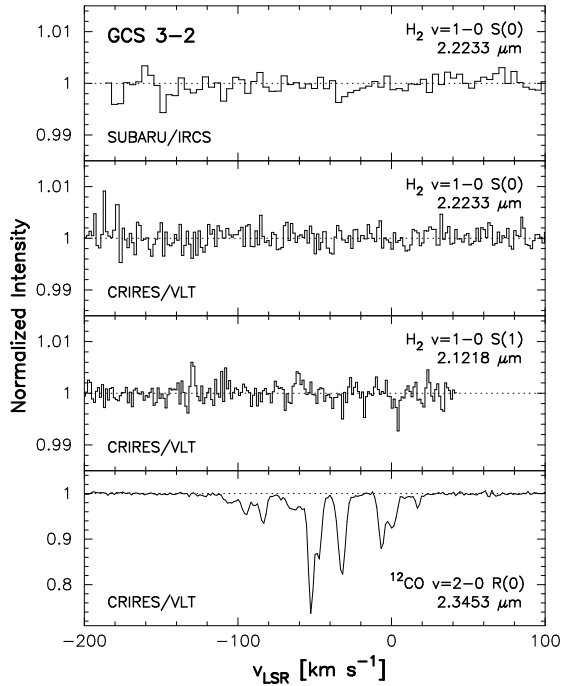}
\caption{Spectra of GCS~3-2 at wavelengths of H$_2$ in the
  2~$\mu$m region. The spectrum of the CO $v$=2-0 $R$(0) line is
  shown for comparison.
\label{h}}
\end{figure}

Figure~\ref{h} shows the spectra of GCS~3-2 obtained at the
wavelengths of the H$_2$ $v$=1-0 $S$(0) and $S$(1) lines, which
in absorption originate from the two lowest lying rotational
levels of the molecule. Neither line was detected.  The radial
velocities and widths of the expected absorption features are
uncertain. From the spectrum of CO (bottom of the figure) one
may assume that H$_{2}$ in foreground arms would produce three
narrow absorptions, each of velocity width
$\sim$6~km~s$^{-1}$. The upper limits on the equivalent widths
of such absorption features in the observed spectrum are
2.3$\times$10$^{-7}$~$\mu$m for each.  For excitation
temperatures of H$_2$ in the range 10--50~K typical of
interstellar clouds, only the $J$=0 level should be
significantly populated. The corresponding upper limit on the
H$_2$ column density is $N$(H$_{2}$)~$<$~
7~$\times$~10$^{21}$~cm$^{-2}$ for each arm. For a standard
dust-to-gas ratio \citep{boh78} and gaseous hydrogen fully
molecular and not depleted by adsorption on grains each limit
corresponds to $A_V$~$<$~7~mag. The accuracies of the
  limits to the extinction derived from the H$_2$ measurements
  are highly uncertain, as are any implications, because the
  fractions of gas in the intervening arms that are in diffuse
  and dense clouds are not well constrained.

\section{Analysis and Discussion}

As described in the previous sections, the observed H$_3^+$ and CO 
absorption lines toward GCIRS~3 and GCIRS~1W are most simply interpreted 
as arising in three distinct physical environments and thus at least 
three distinct locations along those lines of sight. The environments 
and likely locations are: (1) cold and mostly dense clouds in the 
foreground spiral and lateral arms, containing both species; (2) warm 
and diffuse gas within the CMZ, containing H$_3^+$ and relatively little 
CO (undetected in the overtone band of $^{12}$CO nor in the fundamental 
band of $^{13}$CO); and (3) warm and dense compact clouds in the central 
few parsecs of the CMZ, containing both species.  In the first 
three subsections we use the new data and previous work by us and others 
to constrain the physical conditions in each of these environments.  In 
the fourth subsection we briefly discuss possible explanations for the 
enhanced ionization rates in the CMZ's extended diffuse gas and in the 
dense gas within the central few parsecs.

\subsection{Foreground Gas}

The spiral arm local to the solar neighborhood, and possibly
other spiral arms between the sun and the Galactic center are
likely contributors to the strong absorption by CO and H$_3^+$
near 0~km~s$^{-1}$. The sharp absorption feature at
$-$53~km~s$^{-1}$ in both species are readily identifiable as
originating in the 3~kpc arm, which was first recognized as
  a stream of neutral clouds by \citet{woe57} in the \ion{H}{1}
  21~cm line and then located radially by \citet{oor58} from the
  tangential point of the stream at longitude of 303\degr~
  \citep[For a recent image of it see][]{dam08}. Likewise, the
absorption feature at $-$32~km~s$^{-1}$ arises in the 4~kpc arm,
first identified by \citet{men70} via the \ion{H}{1} line.

The weak absorption feature in CO at $-$72~km~s$^{-1}$ is close
in velocity to an H~I absorption feature at $-75$~km~s$^{-1}$
observed toward Sgr A by \citet{lis83}, who did not identify the
feature with any previously known cloud.  A similar kinematic
component was recorded in OH 1667-MHz absorption spectrum at
$-80$~km~s$^{-1}$ by \citet{san70} on the same line of sight. It
is unclear if the CO feature observed here is related to either
of the above. Its presence only in low-$J$ levels, its small
angular size, and the lack of absorption features at that
velocity in the H$_3^+$ line profiles suggest that it does not
arise within the CMZ, but rather in a cold, compact, and
presumably dense foreground cloud.

\subsection{Warm and diffuse gas in the CMZ}

\subsubsection{Velocities, Densities and Temperatures}

Warm and diffuse CMZ gas was discovered by \citet{got02} and
characterized by \citet{oka05}, who identified it as a major
constituent of the CMZ. It produces a broad and shallow swath of
absorption in the $R$(1,1)$^l$ and $R$(3,3)$^l$ lines of
H$_3^+$, but no absorption in the $^{12}$CO $v$=2-0 or $^{13}$CO
$v$=1-0 lines. It is found almost entirely at negative
velocities (from $-$180~km~s$^{-1}$ to $+$20~km~s$^{-1}$); thus
the gas producing it is moving outward from the center. The
strength of this absorption, its presence toward stars located
from the Central Cluster to as far east as the Quintuplet
Cluster \citep{oka05,got08}, and its velocity breadth suggest
not only that it extends across all sightlines between the two
clusters, but also that its column length is a significant
fraction of the $\sim$200~pc radius of the CMZ.

The highest negative velocities observed in H$_3^+$, those in
excess of $-$100~km~s$^{-1}$, have also been observed in the
absorption and emission lines at radio wavelengths of several
molecular species, e.g., CO and CS \citep{bal87}. It seems
likely that the H$_3^+$ and the other molecules seen at these
high velocities are physically associated. The radio lines have
often been interpreted as arising near the outer edge of the CMZ
\citep{kai72,sco72,sof95}, in a shell or ring-like structure
\citep[see][for an alternative explanation]{bin91}. If the high
velocity H$_3^+$ absorption also arises there, then H$_3^+$
absorption at lower negative velocities forms interior to
it. However, the lower negative velocity gas must still be
distant radially from the central few tens of parsecs, as
absorption by it is observed on all sightlines from the Central
Cluster to the Quintuplet Cluster.

\begin{figure}
\plotone{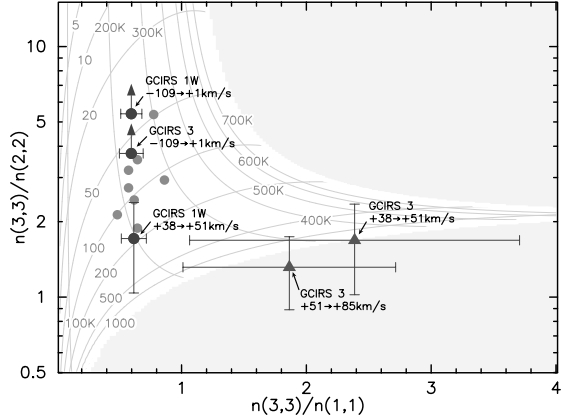}
\caption{Temperature and density diagram for H$_3^+$, with
  $n(3,3)/n(2,2)$ plotted against $n(3,3)/n(1,1)$. Dark circles
  for GCIRS~1W and GCIRS~3 in the velocity intervals $-$109 to
  $+$1~km~s$^{-1}$ and for GCIRS~1W for $+$38 to
  $+$51~km~s$^{-1}$ correspond to the warm and diffuse gas in
  the CMZ. GCIRS~1W at $+$38 to $+$51~km~s$^{-1}$ is shown
  separately because that velocity interval corresponds to a
  detection of weak $R$(2,2)$^l$ line  absorption;
  $n(3,3)/n(2,2)$ for all other points are lower limits. Gray
  circles denote lower limits on sightlines within 30~pc of
  Sgr~A* from \citet{got08}. Filled triangles for GCIRS~3 at
  $+$38 to $+$51~km~s$^{-1}$ and $+$51 to $+$85~km~s$^{-1}$
  correspond to warm, dense and compact clouds probably
  associated with the CND (see text).
  \label{nt}}
\end{figure}

The temperature and density of the diffuse gas can be determined
from the H$_3^+$ level column density ratios
  $N$(3,3)/$N$(1,1) and $N$(3,3)/$N$(2,2) \citep{oka04}. At low
densities the former ratio is mainly temperature dependent and
the latter is mainly density dependent.  Fig.~\ref{nt} shows
temperature and density as functions of these ratios. In the
velocity range $-$109~km~s$^{-1}$ $\rightarrow$$+$1~km~s$^{-1}$
the 1$\sigma$ lower limits of $n$(3,3)/$n$(2,2) toward the two
Central Cluster sources are 3.7 and 5.4, which correspond to
mean temperatures near 250~K and mean densities
$n$~$\lesssim$50~cm$^{-3}$.

In Fig.~\ref{f1} the spectra of GCIRS~1W show warm and
  diffuse gas at positive velocities as high as
  $+$50~km~s$^{-1}$ as evidenced by the weakness of the
  absorption in the $R$(2,2)$^l$ line. This positive velocity
  diffuse gas, previously seen by \citet{got08}, is not clearly
  present on any other Galactic center sightlines observed to
  date, although it may contribute to the absorption profiles
  seen toward GCIRS~3. Fig.~\ref{nt} shows that while this gas
  is diffuse it is somewhat higher density than the diffuse gas
  on other GC sightlines.

\subsubsection{Ionization Rate} \label{4.2.2}

The simple chemistry of H$_3^+$ in the diffuse interstellar
medium allows one to determine the product $\zeta$$L$ from the
equation \citep{oka05,oka06}

\begin{equation}
  \zeta L  = 2 k_{e} N({\rm H_3^+})_{\rm total}(n_{\rm C}/n_{\rm H})_{\rm SV}R_{X}/f({\rm H_2}), 
\end{equation}

\noindent where $k_{e}$ is the rate constant for the
dissociative recombination of H$_3^+$ on electrons, ($n_{\rm
  C}$/$n_{\rm H}$)$_{\rm SV}$ is the carbon to hydrogen ratio in
diffuse clouds in the solar vicinity, $R_{X}$ is the factor
increase of that ratio from the solar vicinity to the GC (due to
higher metallicity in the GC), and
$f$(H$_2$)~=~2$n$(H$_2$)/$n_{\rm H}$ is the fraction of hydrogen
in molecular form. We use the value of $k_{e}$ at 230~K,
8.1~$\times$10$^{-8}$~cm$^3$~s$^{-1}$, calculated from Eq. (7)
of \citet{mcc04} (note that the recent experiment by
\citet{pet11} implies a somewhat larger value), and ($n_{\rm
  C}$/$n_{\rm H}$)$_{\rm SV}$~=~1.6~$\times$~10$^{-4}$
\citep{sof04}. We use $R$$_{X}$~=~3, which appears to be
conservative lower limit \citep{sod95,ari96,rol00,chi01,est05},
and $f$(H$_2$)~=~1. Then

\[
\zeta L > (7.8\times10^{-11}\,{\rm cm^3\,s^{-1}})\,N({\rm H_3^+})_{\rm total}.
\]

The total H$_3^+$ column densities in the warm and diffuse gas, calculated 
from the sum of $N$(1,1) and $N$(3,3) from $-$180~km~s$^{-1}$ to 
$+$20~km~s$^{-1}$ in Table~\ref{tb2} and $N$(1,0) in Table~4 of 
\citet{got08} (after multiplying by 0.83 in the case of GCIRS~1W to 
adjust for the different velocity intervals), are 
3.0~$\times$10$^{15}$~cm$^{-2}$ toward each source. The actual value may 
be several percent higher because the higher metastable levels such as 
(4,4), (5,5), (6,6) \citep{oka04} and some unstable levels such as 
(2,2), and (2,1) may have non-negligible populations.

We thus obtain $\zeta$$L$~$>$~2.3~$\times$~10$^5$~cm~s$^{-1}$,
which is similar to the lower limit found toward GCS~3-2
\citep{oka05} and toward other stars from Sgr~A* to 30~pc east
\citep{got08}. The limit is more than an order of magnitude
higher than values in diffuse clouds in the Galactic disk
(0.5--1.9)~$\times$~10$^4$~cm~s$^{-1}$ \citep{mcc02,ind07} and
more than three orders of magnitude higher than values in
Galactic dense clouds \citep{mcc99}. This may be partly
  attributable to the line of sight extent of diffuse gas in the
  CMZ being considerably longer than in Galactic diffuse
  clouds. However, use of the average value of $\zeta$ in
Galactic diffuse clouds, 3.5~$\times$~10$^{-16}$~s$^{-1}$
\citep{ind12}, gives $L~>$~200~pc, which is comparable to the
radius of the CMZ.  Since it is unlikely that the warm and
  diffuse gas fills the entirety of the foreground CMZ, $\zeta$
  is probably considerably greater than its average value in
Galactic diffuse clouds.

This conclusion is strengthened by the likelihood that $f$(H$_2$) is 
considerably less than unity in the warm and diffuse gas in the CMZ. No 
information about $f$(H$_2$) on Central Cluster sightlines is available. 
Elsewhere in the CMZ recent Herschel observations have revealed strong 
H$_2$O$^+$ line emission towards Sgr~B2 \citep{sch10} with a complex 
velocity profile strikingly similar to that found for H$_3^+$ toward 
2MASS~J17470898-2829561 \citep{geb10}, at a projected distance of only 
17~pc from Sgr B2. This similarity indicates that the two sightlines 
pass through the same clouds. Models suggest that H$_2$O$^+$ can only 
exist in regions where $f$(H$_2$) is less than 0.1, since H$_2$O$^+$ 
would be quickly destroyed through the reaction H$_2$O$^+$ + H$_2$ 
$\rightarrow$ H$_3$O$^+$ \citep{ger10}.  If $f$(H$_2$) smaller than 
0.1 is appropriate for the sightlines toward the Central Cluster and 
Quintuplet sources the ionization rates there may exceed 
10$^{-14}$~s$^{-1}$. It would be surprising if such large 
amounts of H$_3^+$ as observed in the CMZ can be generated if $f$(H$_2$) 
is that small. It does seem safe to conclude that 
$\zeta$~$>$~1$\times$~10$^{-15}$~s$^{-1}$ in the CMZ's diffuse gas.

\subsection{Warm, Dense and Compact Clouds} \label{4.3}

At positive velocities the spectra of H$_3^+$ and $^{13}$CO
reveal the presence of two warm and dense clouds. The most
redshifted of these, observed only toward GCIRS~3, produces
broad and asymmetric absorption features in both $^{13}$CO and
H$_3^+$. In $^{13}$CO the feature extends from $+$50~km~s$^{-1}$
to +85~km~s$^{-1}$, with maximum absorption near
$+$60~km~s$^{-1}$ and is observable out to $J$=6
(Fig.~\ref{co1}). In H$_3^+$ a similarly asymmetric absorption
feature extends from $+$40~km~s$^{-1}$ to $+$80~km~s$^{-1}$,
with maximum absorption near $+$55~km~s$^{-1}$, in all
  three lines including $R$(2,2)$^l$ (Fig.~\ref{f1}).

The other cloud produces a narrow and weak absorption
(FWHM$\sim$10~km~s$^{-1}$) centered near $+$45~km~s$^{-1}$ in
$^{13}$CO toward both GCIRS~1W out to $J$=5 and GCIRS~3, and is
detectable out to $J$=6. In the lower $J$ levels the absorptions
produced by this cloud are stronger toward GCIRS~1W than toward
GCIRS~3. The presence of this cloud as a distinct entity is less
obvious in the spectra of H$_3^+$, which show a broad absorption
centered near $+$40~km~s$^{-1}$ toward GCIRS~1 but
no clear feature toward GCIRS~3. Little or no absorption in
  the H$_3^+$ $R$(2,2)$^l$ line is evident toward GCIRS~1W. In
  the following sections we refer to the clouds producing these
  two velocity features as the $+$60~km~s$^{-1}$ cloud and the
  $+$45~km~s$^{-1}$ cloud.

\subsubsection{Densities and Temperatures} \label{4.3.1}

To estimate physical conditions in the $+$60~km~s$^{-1}$ and
$+$45~km~s$^{-1}$ clouds  toward GCIRS~3, we again use the
H$_3^+$ line ratios, after integrating the H$_3^+$ line
strengths over the velocity ranges of the two clouds. Here
  we have subtracted the spectra of H$_3^+$ toward GCIRS~1W from
  those toward GCIRS~3, under the assumption that the warm
  diffuse gas seen toward the former source extends across the
  latter. The results are shown in Fig.~\ref{nt}.  For the
$+$60~km~s$^{-1}$ cloud ($+$51 to $+$85~km~s$^{-1}$) the data
point is in a region where the inversion of the original diagram
in \citet{oka04} does not have a solution. We use the original
diagram to obtain $T$~=~(300~$\pm$~50)~K and
$n$~$\geq$~10$^4$~cm$^{-3}$. Similar results are obtained for
the $+$45~km~s$^{-1}$ cloud ($+$38 to $+$51~km~s$^{-1}$) at
GCIRS~3.

\input{t7.tex} 

 Excitation temperatures and total $^{13}$CO column
  densities in the positive velocity clouds toward GCIRS~3 and
  GCIRS~1W were estimated from their population diagrams in
  Fig.~\ref{pp}. The results are summarized in Table~\ref{t7}. A
  two-temperature model was used for the $+$45~km\,s$^{-1}$
  cloud toward GCIRS~3 ($+$38 to $+$51~km~s$^{-1}$ in
  Fig.~\ref{pp}), as the level population turns over near
  $J=3$. Although it is likely that the CO is subthermally
  excited it was not possible to reproduce the observed turnover
  in the level populations using the radiation transfer code
  RADEX \citep{tak07}. We note that the critical density for
  collisionally maintaining an LTE population of CO up to $J$=5
  is $\sim$4~$\times$~10$^5$~cm$^{-3}$ \citep{kra04}. Because
  these clouds likely exist in an intense infrared radiation
  field, which might assist in populating the lower rotational
  levels of CO, their effective ``critical" densities could be
  somewhat less. In any event the estimates are consistent with
  those based on the H$_3^+$ lines.

\begin{figure}
\plotone{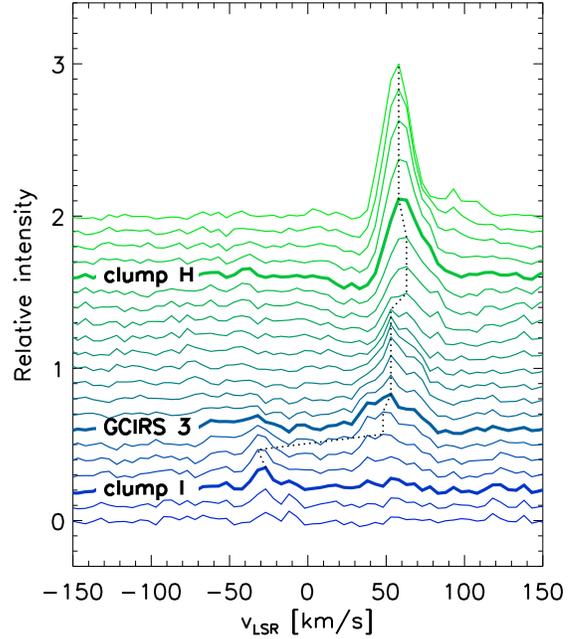}
\caption{HCN $J=$4-3 emission lines \citep{mon09} extracted from
  the apertures between clumps H and I shown in
  Fig.~\ref{hcn_image}. The velocity of the peak line emission
  is marked by a dotted line \citep[originally published
      in][]{got13}.} \label{hcn_line}
\end{figure}

\subsubsection{Location}

The velocity of the $+$60~km~s$^{-1}$ cloud naively suggests
that it may be associated with the well-known giant molecular
cloud M$-$0.02$-$0.07, also known as the ``$+$50~km~s$^{-1}$
cloud'' \citep{bro84}. However, that cloud, which gives the
strongest CO radio emission on this sightline \citep{oka98}, is
located behind the Central Cluster, as surmised from the lack of
$^{12}$CO $v$=1-0 absorption toward GCIRS~1W \citep{geb89} as
well as from radio studies of NH$_{3}$ \citep[][see their
  Fig.~1]{lho91,coi00}.  Both the close proximity of GCIRS~3 to
Sgr~A* in the plane of the sky, suggesting that GCIRS~3
  resides in the central parsec, and the requirement of a sharp
edge in the cloud to provide absorption toward GCIRS~3 but not
toward GCIRS~1W argue against the absorber being
M$-$0.02$-$0.07.

\begin{figure}
\plotone{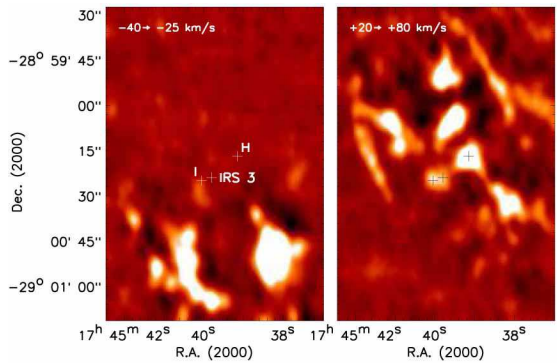}
\caption{HCN~$J$=4-3 maps \citep{mon09} integrated over the
  velocity intervals $-$40 to $-$25~km~s$^{-1}$ (left) and $+$20
  to $+$80~km~s$^{-1}$ (right). The locations of clump H, I, and
  GCIRS~3 are shown by crosses.
\label{hcn_cross}}
\end{figure}

We suggest instead that the $+$60~km~s$^{-1}$ cloud is
associated with the Galactic center's Circumnuclear Disk (CND),
as originally proposed by \citet{geb89}. The locations of the
stars in the Central Cluster are compared to an HCN~$J$=4-3 map
of \cite{mon09} in Fig.~\ref{hcn_image}. GCIRS~3 coincides with
an extension of the northwest portion of the CND to the
east-southeast (i.e. into the interior of the CND) termed
``clump~I'' by \citet{mon09}, while the line of sight to
GCIRS~1W is clear. Clump I is close to Sgr~A*, and thus far from
the inner edge of the CND. In order to determine if this
extension is related to the CND, HCN~$J$=4-3 spectra were
extracted along the line connecting clump H (in the main portion
of the CND as seen in Fig.~\ref{hcn_image}) and clump I. The
spectra are shown in Fig.~\ref{hcn_line}. The radial velocity at
the line peak smoothly changes from $+$60~km~s$^{-1}$ to
$+$50~km~s$^{-1}$ from clump~H to the position of GCIRS~3 at the
western extension of clump~I, as well as steadily decreasing in
intensity. Emission at this velocity disappears just beyond
clump I; the intensity peak there is due to emission at
$-$30~km~s$^{-1}$. This shift in radial velocity of the peak is
also visible in the velocity-integrated map of HCN~$J$=4-3 in
Fig.~\ref{hcn_cross}. Thus the western extension of clump~I,
which intersects the line of sight to GCIRS~3, is physically
connected to clump~H of the CND, rather than to the source of
emission peak at clump~I at $-$30~km~s$^{-1}$, which likely is
situated in the background.

\begin{figure}
\plotone{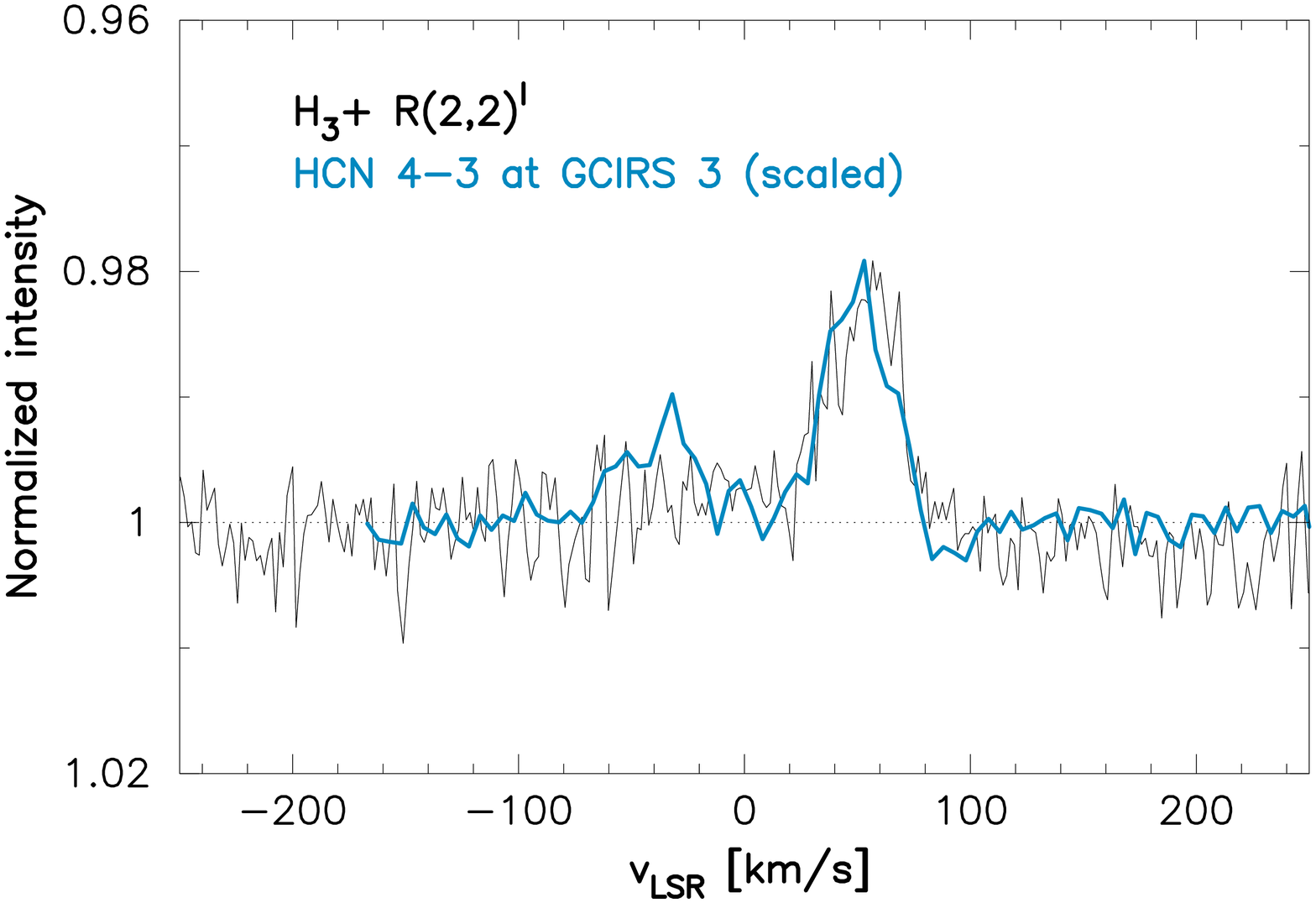}
\caption{Comparison of H$_3^+$~$R$(2,2)$^l$ absorption toward
  GCIRS~3 (black) to HCN~$J$=4-3 emission line \citep{mon09}
  extracted at the location of GCIRS~3 (blue, scaled to
  $R$(2,2)$^l$)  \citep[originally published
      in][]{got13}.
\label{hcn_r2}}
\end{figure}

The H$_3^+$ $R$(2,2)$^l$ absorption line toward GCIRS~3 is compared to 
HCN~$J$=4-3 emission line extracted at the position of GCIRS~3 in 
Fig.~\ref{hcn_r2}. The line profiles match almost perfectly in terms of 
their line center velocities and widths. We conclude that the 
$+$60~km~s$^{-1}$ absorption observed toward GCIRS~3 in $^{13}$CO and 
H$_3^+$ occurs in the western extension of clump~I, which is a part of 
the CND. The extension was not resolved by the 4\farcs6$\times$3\farcs0 
synthesized beam of the SMA \citep{mon09}; thus its linear dimension is 
less than 0.2~pc. 

 It is not surprising that this gas, observed by us up to
  $J$=6 in $^{13}$CO, is warm. The presence of high temperature
  gas in the CND, although not specifically on this sightline,
  has been known from millimeter wave spectroscopy $^{12}$CO
  populated up to $J=7$ \citep{mon01,bra05}, and up to $J=24$
  \citep{goi13}).  \citet{bra05}, \citet{req12}, and
  \citet{goi13} compared multiple CO rotational lines with
  non-LTE radiation transfer models to estimate the ambient gas
  density to be $10^4$--$10^5$~cm$^{-3}$. Previously
  \citet{chr05} and \citet{mon09} estimated the gas density of
  the clumps in their HCN maps to be $>10^8$~cm$^{-3}$, assuming
  the clumps are virialized. \citet{mil13} recently argued,
  however, that in the CND the effective critical density of HCN
  is reduced by radiative pumping in the mid-infrared in the
  CND, $\sim$$10^4$--$10^5$~cm$^{-3}$. Our H$_3^+$ spectroscopy
  seems to be more consistent with the the latter values.

The mass of the $+$60~km~s$^{-1}$ cloud may be crudely estimated
assuming a diameter given by the separation of GCIRS~1W and
GCIRS~3 and the gas density derived from H$_3^+$ spectroscopy,
$\sim$$10^4$~cm$^{-3}$. We obtain $M$~$\sim$~4~$M_{\odot}$.  We
are unable to identify the material producing the
$+$45~km~s$^{-1}$ feature toward GCIRS~1W and GCIRS~3 with any
known cloud and do not attempt to estimate its mass.

\subsubsection{Ionization Rate}

For dense clouds, in which electrons are scarce and proton hop reactions 
from H$_3^+$ to CO are the main destruction channel for H$_3^+$, we use 
the analog to Equation 1,

\begin{equation}
\zeta L = k_{L} N ({\rm H_3^+})_{\rm total}(n_{\rm CO}/n_{\rm H_2})_{\rm SV}R_{X},
\end{equation}

\noindent where the Langevin rate constant for CO, 
$k_{L}$=~2~$\times$~10$^{-9}$~cm$^3$~s$^{-1}$ \citep{ani86}, replaces 
$k_{e}$ in Equation 1. For the CO to H$_2$ ratio we use 
8~$\times$~10$^{-4}$, the value measured by in the dense cloud in front 
of NGC~2024~IRS~2 by \citet{lac94}, and then multiply by the lower limit 
$R_X=3$ in view of the higher metallicity in the GC, as discussed 
earlier. Using the observed total H$_3^+$ column density for the 
$+$60~km~s$^{-1}$ cloud toward GCIRS~3, 6~$\times$~10$^{14}$~cm$^{-2}$ 
(the sum of the column densities in the bottom two rows in Table~2), and 
multiplying by 1.5 to account for destruction of H$_3^+$ by other dense 
cloud species, most notably O, we obtain 
$\zeta$$L$~$>$~1.5~$\times$~10$^3$~cm~s$^{-1}$. If the path length 
through the $+$60~km~s$^{-1}$ cloud is 0.3~pc, 
$\zeta$~$>$~1.6~$\times$~10$^{-15}$~s$^{-1}$. As in the case of the 
diffuse CMZ gas, this lower limit is significantly greater than mean 
values for dense or diffuse clouds outside of the GC.

\subsection{High Ionization Rates in the CMZ}

The sources of the enhanced ionization rates in the CMZ
  ($\zeta$~$>$~1$\times$~10$^{-15}$~s$^{-1}$ found here and also
  by \citet{oka05} and \citet{got08} in the widely distributed
  diffuse gas, and $\zeta$~$>$~1.6~$\times$~10$^{-15}$~s$^{-1}$
  found here for the dense $+$60~km~s$^{-1}$ cloud associated
  with the CND are not clearly identified. Both increased cosmic
  ray particle fluxes, arising from the relatively high
  concentration of supernova remnants in the GC, and enhanced
  photoionization by ultraviolet and X-ray photons from the the
  nearby hot and luminous stars, supernova remnants, accretion
  disks of black holes, and ultra-hot diffuse gas are possible
  contributors \citep{cro11,mun09,yus07a}.

In the CMZ ionization of hydrogen due to UV radiation would
  be strongly influenced by the distribution of gas and dust.
  For gas densities of 100~cm$^{-3}$ or more, mean free paths
  for UV photons are small fractions of a parsec. The situation
  is different for X-rays. The X-ray photoionization cross
  section per hydrogen atom is $\sim 10^{-22}$~cm$^2$ per
  hydrogen atom at 1~keV \citep{wil02}. In diffuse gas the mean
  free path of a 1~keV X-ray photon is a few tens of
  parsecs. X-ray images of the Galactic center reveal an
  extended diffuse component and a plethora of point sources
  with no individual source or group of sources that are
  dominant \citep{bag03,mun09}.  Thus it would not be surprising
  if the ionization rate, even if enhanced due to X-rays, or due
  to cosmic ray particles, were roughly constant on widely
  spaced sightlines through the extended warm and diffuse gas in
  the CMZ, as seems to be the case \citep{oka05,got08,got11}.

On the other hand, the high ionization rate measured in the 
$+$60~km~s$^{-1}$ cloud, which is located close to the Central Cluster 
and to Sgr A*, might be attributed in part to recent flares of local 
X-ray sources such as Sgr A* itself, or possibly to UV ionization from 
the multitude of hot stars in the nearby Central Cluster. For more 
detailed discussion of ionization rates in the central few pc of the 
Galaxy, see \citet{got13}, who estimated X-ray and cosmic ray ionization 
rates based on X-ray and $\gamma$-ray observations near Sgr~A*.

\section{Conclusion}

The basic physical properties of the gas in the Central Molecular Zone 
of the Galaxy have been quantified on sightlines toward infrared stars 
in the Central Cluster that are within a few tenths of a parsec of Sgr 
A*, using new infrared absorption spectra of H$_3^+$ and CO.  
Two types of gaseous environments within the CMZ have been identified on 
these sightlines: (1) warm (200~K~$<$~$T$~$<$~300~K) and diffuse 
($n$~$\leq$~100~cm$^{-3}$) gas with velocities in the range 
$-$180~km~s$^{-1}$ to $+$20~km~s$^{-1}$ occupying a significant 
fraction of the outer portion of the CMZ; (2) warm ($T$~$\sim$~300~K) 
and dense ($n$~$\geq$10$^4$~cm$^{-3}$) gas probably belonging to an 
inward extension of the 1.5~pc radius Circumnuclear Disk that is on the 
line of sight to GCIRS~3. In addition, cold dense and diffuse gas 
located in foreground spiral and lateral arms has been observed on these 
sightlines.

From the observed total column densities of H$_3^+$, products of 
ionization rate $\zeta$ and pathlength $L$ have been determined to be 
$\zeta$$L$~$>$~2.3~$\times$~10$^5$~cm~s$^{-1}$ and 
$\zeta$$L$~$>$~1.5~$\times$~10$^3$~cm~s$^{-1}$, respectively, for the 
above diffuse and dense CMZ gas. Although separation of $\zeta$ and $L$ 
is difficult, the large values of their products indicate ionization 
rates ($\zeta > $10$^{-15}$~s$^{-1}$) in both environments, and large 
path lengths ($L>$30~pc) in the diffuse gas of the CMZ.

 \acknowledgments {We are grateful to the staffs of the VLT and
   the Subaru Telescope for valuable assistance in obtaining
   these data.  We thank the anonymous referee for constructive
   suggestions that markedly improved the paper. We also thank
   Mar{\'i}a Montero-Casta{\~n}o for providing HCN~$J=$4-3 map
   in machine-readable form. This research has made use of the
   SIMBAD database, operated at CDS, Strasbourg, France. The
   image of the Central Cluster was obtained through the ESO
   Science Archive Facility. T. O. is supported by NSF grant AST
   1109014.  M. G. is supported by DFG grant
   GO~1927/3-1. T. R. G.'s research is supported by the Gemini
   Observatory, which is operated by the Association of
   Universities for Research in Astronomy, Inc., on behalf of
   the international Gemini partnership of Argentina, Australia,
   Brazil, Canada, Chile, the United Kingdom, and the United
   States of America. We appreciate the hospitality of Chilean
   and Hawaiian communities, who made this research possible.}

\end{document}

%% file: t1.tex
\begin{deluxetable*}{ll c l r l c cc}
\tabletypesize{\footnotesize}
\tablewidth{0pt}
\tablecaption{Summary of Observations.\label{t1}}
\tablehead{
\colhead{Object}           & 
\colhead{UT Date}          &        
\colhead{Instrument/Telescope}          &        
\colhead{Lines}            &
\colhead{Grating \tablenotemark{a}} &
\colhead{Spec. Res.}        & 
\colhead{Exp.\tablenotemark{b}} &
\colhead{Standard}} 

\startdata

GCIRS~3/GCIRS~1W    & 11 Oct 2006         &CRIRES/VLT  & $^{13}$CO $v$=1-0    & 12/1/n  & $R$=50,000  & 1   & HR~6879 \\
GCIRS~3/GCIRS~1W    & 11 Oct 2006         &CRIRES/VLT  & $^{13}$CO $v$=1-0    & 12/1/i  & $R$=50,000  & 1   & HR~6879 \\
GCIRS~3/GCIRS~1W    & 9 Jun, 4-5 Aug 2007 &CRIRES/VLT  & H$_3^+$ $R$(1,1)$^l$ & 3739.4  & $R$=100,000 & 108 & HR~6879 \\
GCIRS~3/GCIRS~1W    & 5, 9, 10 Aug 2007   &CRIRES/VLT  & H$_3^+$ $R$(2,2)$^l$ & 3646.1  & $R$=100,000 & 126 & HR~6879 \\
GCIRS~3/GCIRS~1W    &10, 28 Aug 2007      &CRIRES/VLT  & H$_3^+$ $R$(3,3)$^l$ & 3533.6  & $R$=100,000 & 108 & HR~6879 \\
GCIRS~16NE/GCIRS~21 & 13 May 2007         &CRIRES/VLT & $^{12}$CO $v$=2-0     & 2239.2  & $R$=100,000 & 40  & HR~6879 \\
GCIRS~16NE/GCIRS~21 & 14 May 2007         &CRIRES/VLT & $^{12}$CO $v$=2-0     & 2336.2  & $R$=100,000 & 40  & HR~6879 \\

GCS~3-2             & 15 Sep 2007         &CRIRES/VLT & $^{12}$CO $v$=2-0     & 2239.2  & $R$=100,000 & 16  & HR~6879 \\
GCS~3-2             & 15 Sep 2007         &CRIRES/VLT & $^{12}$CO $v$=2-0     & 2336.2  & $R$=100,000 & 16  & HR~6879 \\

GCS~3-2             & 27 Jul 2008         &CRIRES/VLT & H$_2$ $v$=1-0 $S$(0)  & 2236.1  & $R$=100,000 & 80  & HR~6879 \\
GCS~3-2             & 15 Oct 2007         &CRIRES/VLT & H$_2$ $v$=1-0 $S$(1)  & 2117.6  & $R$=100,000 & 40  & HR~6879 \\
GCS~3-2             & 17 Aug 2008         &CRIRES/VLT & H$_2$ $v$=1-0 $S$(1)  & 2117.6  & $R$=100,000 & 40  & HR~6879 \\

\hline
GCS~3-2             & 24 May 2003        & IRCS/SUBARU & H$_2$ $v$=1-0 $S$(0) & 4300/200  & $R$=20,000  &  60 & HR~7557,HR~7194    
\enddata

\tablenotetext{a}{ For CRIRES these are reference wavelength
    in nm, or grating setting name for 2006 observations
    only. For IRCS the entry is echelle and cross-disperser
    angle.}

\tablenotetext{b}{Total integration time in minutes.}

\end{deluxetable*}
\normalsize

%% file: t2.tex
\begin{deluxetable*}{l c ccc ccc}
\tablecolumns{8}
\tablecaption{Equivalent widths $W_\lambda$ and column densities of H$_3^+$ in the CMZ toward GCIRS~3 and GCIRS~1W.\label{tb2}}
\tablehead{
\colhead{Object} &
\colhead{$v_{\rm LSR}$  [km~s$^{-1}$]} &
\multicolumn{3}{c}{$W_\lambda$ [10$^{-6}\mu$m]} &
\multicolumn{3}{c}{$N(J,K)$ [10$^{14}$cm$^{-2}$]} \\
\colhead{}&
\colhead{}&
\multicolumn{3}{c}{\hrulefill}&
\multicolumn{3}{c}{\hrulefill}\\
\colhead{}&
 \colhead{}&
 \colhead{$R$(1,1)$^l$} &
 \colhead{$R$(3,3)$^l$} &
 \colhead{$R$(2,2)$^l$} &
 \colhead{(1,1)} &
 \colhead{(3,3)} &
 \colhead{(2,2)} }
\startdata
GCIRS~3  &$-$180 $\rightarrow$ $-$109                  & 9.43$\pm$1.59  & 7.49$\pm$1.10  & $<$2.51         & 4.33$\pm$0.73        & 2.66$\pm$0.39        &  $<$0.94             \\
         &$-$109 $\rightarrow$ $+$1 \tablenotemark{a}  & 21.3$\pm$2.5   & 16.4$\pm$1.7   & $<$4.17         & 9.75$\pm$1.17        & 5.82$\pm$0.61        &  $<$1.56             \\
         &$+$1   $\rightarrow$ $+$38\tablenotemark{b}  & 2.15$\pm$1.06  & 1.48$\pm$1.01  & $<$1.14         & 0.99$\pm$0.49        & 0.53$\pm$0.36        &  $<$0.43             \\
         &$+$38  $\rightarrow$ $+$51\tablenotemark{b}  & 0.66$\pm$0.34  & 2.02$\pm$0.36  & 1.14$\pm$0.40   & 0.30$\pm$0.16        & 0.72$\pm$0.13        &  0.43$\pm$0.15       \\
         &$+$51 $\rightarrow$  $+$85\tablenotemark{b}  & 2.30$\pm$0.98  & 5.51$\pm$0.93  & 3.98$\pm$1.10   & 1.05$\pm$0.45        & 1.96$\pm$0.33        &  1.49$\pm$0.41       \\

\tableline

GCIRS~1W &$-$180 $\rightarrow$ $-$109                  & 8.91$\pm$1.22  & 6.00$\pm$0.96  & $<$1.67         & 4.09$\pm$0.56        & 2.13$\pm$0.34        &  $<$0.63             \\
         &$-$109 $\rightarrow$ $+$1 \tablenotemark{a}  & 19.2$\pm$1.9   & 14.8$\pm$1.5   & $<$2.62         & 8.82$\pm$0.88        & 5.26$\pm$0.53        &  $<$0.98             \\
         &$+$1   $\rightarrow$ $+$38                   & 6.58$\pm$0.65  & 5.49$\pm$0.50  & 1.04$\pm$0.81   & 3.02$\pm$0.30        & 1.95$\pm$0.18        &  0.39$\pm$0.31       \\
         &$+$38  $\rightarrow$ $+$51                   & 1.71$\pm$0.20  & 1.36$\pm$0.15  & 0.76$\pm$0.28   & 0.79$\pm$0.09        & 0.48$\pm$0.05        &  0.28$\pm$0.11       \\
         &$+$51  $\rightarrow$ $+$64                   & 0.74$\pm$0.19  & 0.46$\pm$0.15  & 0.40$\pm$0.24   & 0.34$\pm$0.09        & 0.16$\pm$0.05        &  0.15$\pm$0.09       


\enddata
\tablenotetext{a}{Equivalent widths of $R$(1,1)$^l$ at the
  velocity range $-$109$\rightarrow$ $+$1~km~s$^{-1}$ are
  calculated by scaling $R$(3,3)$^l$ absorption lines at the
  same velocity interval by the factor of 1.3 so that
  $R$(3,3)$^l$ matches to the trough absorption of $R$(1,1)$^l$,
  in order to disentangle the absorptions in the CMZ from the foreground arm clouds.}

\tablenotetext{b}{Equivalent widths of $R$(1,1)$^l$, $R$(3,3)$^l$, and $R$(2,2)$^l$
toward GCIRS~1W are subtracted to isolate the local 
absorption components of H$_3^+$ to GCIRS~3.}  


\end{deluxetable*}

%% file: t3.tex
\begin{center}
\begin{deluxetable*}{l cccc}

\tablecolumns{5}
\tablewidth{0pt}
\tablecaption{Equivalent widths $W_\lambda$ and column densities of
  H$_3^+$ in the Galactic arm clouds toward GCIRS~3 and toward
  GCIRS~1W.\label{tb3}}

\tablehead{
\colhead{Object}&
\colhead{$v_{\rm LSR}$ [km~s$^{-1}$]}&
\colhead{$W_\lambda$\tablenotemark{a} [10$^{-6}\mu$m]}&
\colhead{$N(J,K)$\tablenotemark{a}[10$^{14}$cm$^{-2}$]}\\ 
\colhead{}&
\colhead{}&
\colhead{$R$(1,1)$^l$}&
\colhead{(1,1)}}

%

\startdata
GCIRS~3  & $-$62 $\rightarrow$ $-$45    & 2.95$\pm$0.72 & 1.35$\pm$0.33 \\
         & $-$37 $\rightarrow$ $-$25    & 0.78$\pm$0.51 & 0.36$\pm$0.23 \\
         & $-$15 $\rightarrow$ $+$12    & 6.57$\pm$1.17 & 3.01$\pm$0.54 \\

\tableline
GCIRS~1W & $-$62 $\rightarrow$ $-$45    & 3.61$\pm$0.57 & 1.66$\pm$0.26 \\ 
         & $-$37 $\rightarrow$ $-$25    & 1.88$\pm$0.39 & 0.86$\pm$0.18 \\ 
         & $-$15 $\rightarrow$ $+$12    & 5.90$\pm$0.94 & 2.71$\pm$0.43 

\enddata

\tablenotetext{a}{ After removal of CMZ trough component, assumed
  to be given by the $R$(3,3)$^l$ absorption profile scaled by
  1.3.}
\end{deluxetable*}
\end{center}
\normalsize

%% file: t4.tex
\begin{deluxetable*}{l c ccccc ccccc c}

\tablecaption{Equivalent widths $W_\lambda$ of CO.\label{tb4}}

\tablehead{
\colhead{} &
\colhead{$v_{\rm LSR}$ [km~s$^{-1}$]} & 
\multicolumn{11}{c}{$W_\lambda$  [10$^{-5}\mu$m]} \\
\colhead{} & 
\colhead{$^{13}$CO v=1-0}   &
\colhead{$P$(4)}&
\colhead{$P$(3)}&
\colhead{$P$(2)}&
\colhead{$P$(1)}&
\colhead{$R$(0)}&
\colhead{$R$(1)}&
\colhead{$R$(2)}&
\colhead{$R$(3)}&
\colhead{$R$(4)}&
\colhead{$R$(5)}&
\colhead{$R$(6)}}
\startdata
     GCIRS~3&$-$  82$\rightarrow$$-$  65 & ---  & ---  & ---  & 0.86 & 2.86 & 1.18 & 2.24 & ---  & ---  & ---  & ---  \\
            &$-$  65$\rightarrow$$-$  45 & ---  & ---  & ---  & 6.13 &15.07 & 9.94 & 2.45 & ---  & ---  & ---  & ---  \\
            &$-$  37$\rightarrow$$-$  25 & ---  & ---  & ---  & 3.20 &10.39 & 8.28 & ---  & ---  & ---  & ---  & ---  \\
            &$-$  15$\rightarrow$$+$  12 & ---  & 2.08 & 8.61 &18.08 &33.04 & ---  & ---  & 4.41 & ---  & ---  & ---  \\
            &$+$  12$\rightarrow$$+$  38 & ---  & ---  & 5.08 & 7.06 &11.33 &14.33 & 7.48 & 2.61 & ---  & ---  & ---  \\
            &$+$  38$\rightarrow$$+$  51 & ---  & 1.54 & 3.11 & 3.39 & 5.64 & ---  & 6.05 & 2.54 & 1.86 & 1.72 & 1.66 \\
            &$+$  51$\rightarrow$$+$  85 &11.05 &10.28 & 9.60 & 6.93 & ---  & ---  & ---  &14.19 &11.43 & 8.34 & 6.52 \\

\tableline
            &$v_{\rm LSR}$ [km~s$^{-1}$] & \multicolumn{11}{c}{$W_\lambda$  [10$^{-5}\mu$m]} \\
            & $^{13}$CO v=1-0            &$P$(4)&$P$(3)&$P$(2)&$P$(1)&$R$(0)&$R$(1)&$R$(2)&$R$(3)&$R$(4)&$R$(5)&$R$(6)\\
\hline 
    GCIRS~1W&$-$  82$\rightarrow$$-$  65 & ---  & ---  & ---  & ---  & ---  & ---  & ---  & ---  & ---  & ---  & ---  \\
            &$-$  65$\rightarrow$$-$  45 & ---  & ---  & ---  & 6.67 &14.81 & 9.82 & 3.54 & ---  & ---  & ---  & ---  \\
            &$-$  37$\rightarrow$$-$  25 & ---  & ---  & ---  & 2.01 & 8.43 & 5.27 & ---  & ---  & ---  & ---  & ---  \\
            &$-$  15$\rightarrow$$+$  12 & ---  & 1.99 & 9.09 &16.74 &33.13 & ---  & 8.02 & 4.57 & ---  & ---  & ---  \\
            &$+$  12$\rightarrow$$+$  38 & 3.33 & 1.91 & 5.36 & 6.13 & 7.83 &12.79 & 5.04 & 3.29 & ---  & ---  & ---  \\
            &$+$  38$\rightarrow$$+$  51 & ---  & 5.08 & 8.67 & 7.59 & 9.92 &12.53 &14.06 & 6.48 & 2.65 & 1.11 & ---  \\
            &$+$  51$\rightarrow$$+$  64 & ---  & ---  & 1.48 & 2.36 & 2.63 & 3.34 & ---  & 1.70 & 0.56 & ---  & ---  \\

\tableline
            &$v_{\rm LSR}$ [km~s$^{-1}$] & \multicolumn{11}{c}{$W_\lambda$  [10$^{-6}\mu$m]} \\
            & $^{12}$CO v=2-0            &$P$(4)&$P$(3)&$P$(2)&$P$(1)&$R$(0)&$R$(1)&$R$(2)&$R$(3)&$R$(4)&$R$(5)&     \\
\hline 
  GCIRS~16NE&$-$  82$\rightarrow$$-$  65 & ---  & ---  & 1.34 & 2.15 & 2.11 & 2.75 & ---  & ---  & ---  & ---  &      \\
            &$-$  65$\rightarrow$$-$  45 & 1.56 & 1.46 & 8.30 &10.53 &16.20 &16.99 & 6.03 & 1.73 & ---  & ---  &      \\
            &$-$  37$\rightarrow$$-$  25 & 0.89 & ---  & 1.94 & 3.81 & 6.41 & 4.83 & 2.75 & ---  & ---  & ---  &      \\
            &$-$  15$\rightarrow$$+$  12 & 3.45 & 5.10 &10.53 &19.48 &23.97 &25.96 &21.01 & 6.62 & 1.66 & 1.06 &      \\
            &$+$  12$\rightarrow$$+$  38 & ---  & 1.90 & 3.68 & 3.59 & ---  & 4.57 & 8.43 & 4.51 & 1.00 & ---  &      \\
            &$+$  38$\rightarrow$$+$  51 & ---  & 1.52 & 4.17 & 2.03 & 1.04 & 4.63 & 3.41 & ---  & ---  & ---  &      \\
            &$+$  51$\rightarrow$$+$  64 & ---  & 1.00 & 2.75 & 1.47 & 0.94 & 4.03 & 3.21 & ---  & ---  & ---  &      \\

\tableline
            &$v_{\rm LSR}$ [km~s$^{-1}$] & \multicolumn{11}{c}{$W_\lambda$  [10$^{-6}\mu$m]} \\
            & $^{12}$CO v=2-0            &$P$(4)&$P$(3)&$P$(2)&$P$(1)&$R$(0)&$R$(1)&$R$(2)&$R$(3)&$R$(4)&$R$(5)&     \\
\hline 
    GCIRS~21&$-$  82$\rightarrow$$-$  65 & ---  & ---  & ---  & ---  & ---  & ---  & ---  & ---  & ---  & ---  &      \\
            &$-$  65$\rightarrow$$-$  45 & ---  & ---  & 6.45 &11.17 &16.93 &11.73 & 5.49 & ---  & ---  & ---  &      \\
            &$-$  37$\rightarrow$$-$  25 & 2.13 & 2.31 & ---  & ---  & 6.76 & 5.51 & 2.52 & ---  & ---  & ---  &      \\
            &$-$  15$\rightarrow$$+$  12 & 7.76 & 7.83 & 8.96 &22.44 &32.61 &28.06 &19.72 &11.37 & 7.38 & ---  &      \\
            &$+$  12$\rightarrow$$+$  38 & ---  & ---  & ---  & ---  & ---  & ---  & ---  & ---  & ---  & ---  &      \\
            &$+$  38$\rightarrow$$+$  51 & ---  & ---  & 3.39 & 3.29 & ---  & ---  & 3.71 & 2.87 & ---  & ---  &      \\
            &$+$  51$\rightarrow$$+$  64 & ---  & ---  & ---  & ---  & 2.73 & ---  & 3.53 & ---  & ---  & ---  &      \\

\tableline
            &$v_{\rm LSR}$ [km~s$^{-1}$] & \multicolumn{11}{c}{$W_\lambda$  [10$^{-6}\mu$m]} \\
            & $^{12}$CO v=2-0            &     &$P$(3)&$P$(2)&$P$(1)&$R$(0)&$R$(1)&$R$(2)&$R$(3)&$R$(4)&$R$(5)&     \\
\hline 
     GCS~3-2&$-$  98$\rightarrow$$-$  75 &      & 0.84 & 3.15 & 3.06 & 5.50 & 6.96 & 4.25 & 1.70 & 0.50 & ---  &      \\
            &$-$  75$\rightarrow$$-$  65 &      & ---  & ---  & ---  & 1.29 & 1.41 & 0.49 & ---  & ---  & ---  &      \\
            &$-$  65$\rightarrow$$-$  45 &      & ---  & 4.77 & 8.30 &16.73 &17.29 & 6.97 & 0.99 & ---  & ---  &      \\
            &$-$  37$\rightarrow$$-$  25 &      & ---  & ---  & 3.55 & 8.65 & 6.98 & 1.70 & ---  & ---  & ---  &      \\
            &$-$  15$\rightarrow$$+$  12 &      & ---  & 1.69 & 2.81 & 9.18 & 7.66 & 2.74 & ---  & ---  & ---  &      \\
            &$+$  12$\rightarrow$$+$  21 &      & ---  & 0.54 & ---  & 1.04 & 0.87 & 0.55 & ---  & ---  & ---  &      

\enddata
\end{deluxetable*}

%% file: t5.tex
\begin{deluxetable*}{l c ccccc ccccc c}

\tablecaption{Level column densities of CO.\label{tb5}}
\tablehead{ 
\colhead{}  &
\colhead{$v_{\rm LSR}$ [km~s$^{-1}$]} &
\multicolumn{11}{c}{$N_{\rm CO}$ [10$^{15}$cm$^{-2}$]}\\
\colhead{}  & 
\colhead{$^{13}$CO v=1-0 } &
\colhead{$P$(4)}&
\colhead{$P$(3)}&
\colhead{$P$(2)}&
\colhead{$P$(1)}&
\colhead{$R$(0)}&
\colhead{$R$(1)}&
\colhead{$R$(2)}&
\colhead{$R$(3)}&
\colhead{$R$(4)}&
\colhead{$R$(5)}&
\colhead{$R$(6)}}\\

\startdata
     GCIRS~3&$-$  82$\rightarrow$$-$  65 & ---  & ---  & ---  & 1.17 & 1.29 & 0.80 & 1.70 & ---  & ---  & ---  & ---  \\
            &$-$  65$\rightarrow$$-$  45 & ---  & ---  & ---  & 8.30 & 6.83 & 6.77 & 1.86 & ---  & ---  & ---  & ---  \\
            &$-$  37$\rightarrow$$-$  25 & ---  & ---  & ---  & 4.34 & 4.71 & 5.64 & ---  & ---  & ---  & ---  & ---  \\
            &$-$  15$\rightarrow$$+$  12 & ---  & 2.19 & 9.71 &24.50 &14.97 & ---  & ---  & 3.52 & ---  & ---  & ---  \\
            &$+$  12$\rightarrow$$+$  38 & ---  & ---  & 5.72 & 9.57 & 5.13 & 9.76 & 5.67 & 2.08 & ---  & ---  & ---  \\
            &$+$  38$\rightarrow$$+$  51 & ---  & 1.62 & 3.51 & 4.59 & 2.56 & ---  & 4.58 & 2.03 & 1.52 & 1.44 & 1.41 \\
            &$+$  51$\rightarrow$$+$  85 &11.17 &10.80 &10.82 & 9.38 & ---  & ---  & ---  &11.31 & 9.39 & 6.99 & 5.55 \\

\tableline
            &$v_{\rm LSR}$ [km~s$^{-1}$] & \multicolumn{11}{c}{$N_{\rm CO}$ [10$^{15}$cm$^{-2}$]}\\
            & $^{13}$CO v=1-0            &$P$(4)&$P$(3)&$P$(2)&$P$(1)&$R$(0)&$R$(1)&$R$(2)&$R$(3)&$R$(4)&$R$(5)&$R$(6)\\
\hline 
    GCIRS~1W&$-$  82$\rightarrow$$-$  65 & ---  & ---  & ---  & ---  & ---  & ---  & ---  & ---  & ---  & ---  & ---  \\
            &$-$  65$\rightarrow$$-$  45 & ---  & ---  & ---  & 9.03 & 6.71 & 6.69 & 2.69 & ---  & ---  & ---  & ---  \\
            &$-$  37$\rightarrow$$-$  25 & ---  & ---  & ---  & 2.72 & 3.82 & 3.59 & ---  & ---  & ---  & ---  & ---  \\
            &$-$  15$\rightarrow$$+$  12 & ---  & 2.09 &10.25 &22.69 &15.01 & ---  & 6.08 & 3.64 & ---  & ---  & ---  \\
            &$+$  12$\rightarrow$$+$  38 & 3.36 & 2.01 & 6.04 & 8.31 & 3.55 & 8.71 & 3.82 & 2.62 & ---  & ---  & ---  \\
            &$+$  38$\rightarrow$$+$  51 & ---  & 5.34 & 9.78 &10.29 & 4.50 & 8.53 &10.66 & 5.17 & 2.18 & 0.93 & ---  \\
            &$+$  51$\rightarrow$$+$  64 & ---  & ---  & 1.67 & 3.20 & 1.19 & 2.28 & ---  & 1.35 & 0.46 & ---  & ---  \\

\tableline
            &$v_{\rm LSR}$ [km~s$^{-1}$] & \multicolumn{11}{c}{$N_{\rm CO}$ [10$^{17}$cm$^{-2}$]}\\
            & $^{12}$CO v=2-0            &$P$(4)&$P$(3)&$P$(2)&$P$(1)&$R$(0)&$R$(1)&$R$(2)&$R$(3)&$R$(4)&$R$(5)&     \\
\hline 
  GCIRS~16NE&$-$  82$\rightarrow$$-$  65 & ---  & ---  & 0.81 & 1.55 & 0.51 & 0.99 & ---  & ---  & ---  & ---  &      \\
            &$-$  65$\rightarrow$$-$  45 & 0.84 & 0.82 & 4.97 & 7.58 & 3.89 & 6.13 & 2.42 & 0.73 & ---  & ---  &      \\
            &$-$  37$\rightarrow$$-$  25 & 0.48 & ---  & 1.16 & 2.74 & 1.54 & 1.74 & 1.10 & ---  & ---  & ---  &      \\
            &$-$  15$\rightarrow$$+$  12 & 1.86 & 2.85 & 6.30 &14.02 & 5.76 & 9.36 & 8.43 & 2.79 & 0.72 & 0.47 &      \\
            &$+$  12$\rightarrow$$+$  38 & ---  & 1.06 & 2.20 & 2.58 & ---  & 1.65 & 3.38 & 1.90 & 0.43 & ---  &      \\
            &$+$  38$\rightarrow$$+$  51 & ---  & 0.85 & 2.50 & 1.46 & 0.25 & 1.67 & 1.37 & ---  & ---  & ---  &      \\
            &$+$  51$\rightarrow$$+$  64 & ---  & 0.56 & 1.65 & 1.05 & 0.23 & 1.45 & 1.29 & ---  & ---  & ---  &      \\

\tableline
            &$v_{\rm LSR}$ [km~s$^{-1}$] & \multicolumn{11}{c}{$N_{\rm CO}$ [10$^{17}$cm$^{-2}$]}\\
            & $^{12}$CO v=2-0            &$P$(4)&$P$(3)&$P$(2)&$P$(1)&$R$(0)&$R$(1)&$R$(2)&$R$(3)&$R$(4)&$R$(5)&     \\
\hline 
    GCIRS~21&$-$  82$\rightarrow$$-$  65 & ---  & ---  & ---  & ---  & ---  & ---  & ---  & ---  & ---  & ---  &      \\
            &$-$  65$\rightarrow$$-$  45 & ---  & ---  & 3.86 & 8.04 & 4.07 & 4.23 & 2.20 & ---  & ---  & ---  &      \\
            &$-$  37$\rightarrow$$-$  25 & 1.15 & 1.29 & ---  & ---  & 1.62 & 1.99 & 1.01 & ---  & ---  & ---  &      \\
            &$-$  15$\rightarrow$$+$  12 & 4.17 & 4.37 & 5.37 &16.15 & 7.83 &10.12 & 7.91 & 4.79 & 3.20 & ---  &      \\
            &$+$  12$\rightarrow$$+$  38 & ---  & ---  & ---  & ---  & ---  & ---  & ---  & ---  & ---  & ---  &      \\
            &$+$  38$\rightarrow$$+$  51 & ---  & ---  & 2.03 & 2.37 & ---  & ---  & 1.49 & 1.21 & ---  & ---  &      \\
            &$+$  51$\rightarrow$$+$  64 & ---  & ---  & ---  & ---  & 0.66 & ---  & 1.41 & ---  & ---  & ---  &      \\

\tableline
            &$v_{\rm LSR}$ [km~s$^{-1}$] & \multicolumn{11}{c}{$N_{\rm CO}$ [10$^{17}$cm$^{-2}$]}\\
            & $^{12}$CO v=2-0            &     &$P$(3)&$P$(2)&$P$(1)&$R$(0)&$R$(1)&$R$(2)&$R$(3)&$R$(4)&$R$(5)&     \\
\hline 
     GCS~3-2&$-$  98$\rightarrow$$-$  75 &      & 0.47 & 1.89 & 2.20 & 1.32 & 2.51 & 1.70 & 0.72 & 0.22 & ---  &      \\
            &$-$  75$\rightarrow$$-$  65 &      & ---  & ---  & ---  & 0.31 & 0.51 & 0.20 & ---  & ---  & ---  &      \\
            &$-$  65$\rightarrow$$-$  45 &      & ---  & 2.86 & 5.97 & 4.02 & 6.23 & 2.80 & 0.42 & ---  & ---  &      \\
            &$-$  37$\rightarrow$$-$  25 &      & ---  & ---  & 2.55 & 2.08 & 2.52 & 0.68 & ---  & ---  & ---  &      \\
            &$-$  15$\rightarrow$$+$  12 &      & ---  & 1.02 & 2.02 & 2.20 & 2.76 & 1.10 & ---  & ---  & ---  &      \\
            &$+$  12$\rightarrow$$+$  21 &      & ---  & 0.32 & ---  & 0.25 & 0.31 & 0.22 & ---  & ---  & ---  &      

\enddata
\end{deluxetable*}

%% file: t6.tex
\begin{center}
\begin{deluxetable*}{l cc cc cc}
\tablehead{
\colhead{$^{12}$CO} & 
\multicolumn{2}{c}{3~kpc Arm}& 
\multicolumn{2}{c}{4~kpc Arm}&
\multicolumn{2}{c}{0~km~s$^{-1}$ (Spiral arms + GC)} \\
\colhead{}&
\multicolumn{2}{c}{$-$65$\rightarrow$$-$45~km~s$^{-1}$}&
\multicolumn{2}{c}{$-$37$\rightarrow$$-$25~km~s$^{-1}$}&
\multicolumn{2}{c}{$-$15$\rightarrow$$+$12~km~s$^{-1}$} \\
\colhead{}&
\multicolumn{2}{c}{\hrulefill}&
\multicolumn{2}{c}{\hrulefill}&
\multicolumn{2}{c}{\hrulefill} \\
\colhead{}&
\colhead{$N_{\rm CO}$}& 
\colhead{$T_{\rm ex}$}& 
\colhead{$N_{\rm CO}$}& 
\colhead{$T_{\rm ex}$}& 
\colhead{$N_{\rm CO}$}& 
\colhead{$T_{\rm ex}$}\\
\colhead{}& 
\colhead{[10$^{17}$~cm$^{-2}$]}& 
\colhead{[K]}& 
\colhead{[10$^{17}$~cm$^{-2}$]}& 
\colhead{[K]}& 
\colhead{[10$^{17}$~cm$^{-2}$]}& 
\colhead{[K]}}
\startdata
GCIRS~16NE  &    16.6$\pm$     0.9&    11.4$\pm$     0.6&     6.2$\pm$     0.5&    12.3$\pm$     1.0&    30.6$\pm$     1.3&    14.6$\pm$     0.5\\
GCIRS~21    &    13.8$\pm$     1.7&     8.4$\pm$     1.3&     9.5$\pm$     1.7&    19.1$\pm$     3.2&    42.6$\pm$     3.4&    17.4$\pm$     1.2\\
GCS~3-2     &    13.6$\pm$     0.8&     8.4$\pm$     0.4&     5.4$\pm$     0.4&     6.1$\pm$     0.4&     6.2$\pm$     0.9&     7.2$\pm$     0.8\\
\hline
Average     &    14.7$\pm$     0.7&     9.4$\pm$     0.7&     7.0$\pm$     0.6&    12.5$\pm$     3.8&    26.5$\pm$     1.2&    13.1$\pm$     0.7
\enddata
\tablecomments{Uncertainties listed are from fitting only.}
\end{deluxetable*}
\end{center}


%% file: t7.tex

\begin{center}
\begin{deluxetable*}{l cc cc}
\tablecaption{Total column densities of $^{13}$CO and excitation
  temperature of Galactic center clouds at positive velocities
\label{t7}}
\tablehead{
\colhead{}&
\multicolumn{2}{c}{$+$45~km~s$^{-1}$ cloud} &
\multicolumn{2}{c}{$+$60~km~s$^{-1}$ cloud} \\
\colhead{}&
\multicolumn{2}{c}{$+$38$\rightarrow$$+$51~km~s$^{-1}$}&
\multicolumn{2}{c}{$+$51$\rightarrow$$+$85~km~s$^{-1}$}\\
\colhead{}&
\multicolumn{2}{c}{\hrulefill}&
\multicolumn{2}{c}{\hrulefill}\\
\colhead{$^{13}$CO} & 
\colhead{$N_{\rm CO}$} & 
\colhead{$T_{\rm ex}$} & 
\colhead{$N_{\rm CO}$} & 
\colhead{$T_{\rm ex}$} \\
\colhead{} &
\colhead{[10$^{16}$~cm$^{-2}$]}& 
\colhead{[K]}& 
\colhead{[10$^{16}$~cm$^{-2}$]} & 
\colhead{[K]}}
\startdata
GCIRS~3        &     0.8$\pm$ 0.1&     9.8$\pm$     1.1&    ---                &    ---              \\
GCIRS~3 (warm) &     1.0$\pm$ 0.2&    50.8$\pm$     6.4&     6.0$\pm$     0.6  &    52.8$\pm$     5.4\\
GCIRS~1W       &     3.2$\pm$ 0.1&    18.4$\pm$     0.7&  ---  &  --- 
\enddata
\tablecomments{Uncertainties listed are from fitting only.}

\end{deluxetable*}
\end{center}

%% file: ms.bbl
\begin{thebibliography}{}

\bibitem[An et al.(2011)]{an11}
An, D., Ram\'irez, S. V., Sellgren, K., Arendt, R. G., Adwin
Boogert, A. C., Robitaille, T. P., Schultheis, M., et al. 2011,
\apj, 736, 133


\bibitem[Anicich \& Huntress (1986)]{ani86} Anicich, V. G., \&
  Huntress, W. T., Jy. 1986, \apjs, 62, 553

\bibitem[Arimoto et al.(1996)]{ari96} Arimoto, N., Sofue,
  Y., \& Tsujimoto, T. 1996, \pasj, 48, 275

\bibitem[Baganoff et al.(2003)]{bag03}
Baganoff, F. K., Maeda, Y., Morris, M., Bautz, M. W., Brandt, W. N.,
Cui, W., Doty, J. P., Feigelson, E. D., Garmire, G. P., Pravdo,
S. H., Ricker, G. R., \& Townsley, L. K. 2003, \apj, 591, 891


\bibitem[Bally et al.(1987)]{bal87}
  Bally, J., Stark, A. A., Wilson, R. W., \& Henkel, C. 1987,
  \apjs, 65, 13

\bibitem[Bartko et al.(2010)]{bar10} 
  Bartko, H., Martins, F., Trippe, S., Fritz, T. K., Genzel, R.,
  Ott, T., Eisenhauer, F., Gillessen, S., Paumard, T., Alexander, T.,
  Dodds-Eden, K., Gerhard, O., Levin, Y., Mascetti, L., Nayakshin, S.,
  Perets, H. B., Perrin, G., Pfuhl, O., Reid, M. J., Rouan, D., Zilka, M., \&
  Sternberg, A. 2010, \apj, 708, 834

\bibitem[Binney et al.(1991)]{bin91}
Binney, J., Gerhard, O. E., Stark, A. A., Bally, J., \& Uchida, K. I.
1991, \mnras, 252, 210

\bibitem[Bohlin, Savage, \& Drake(1978)]{boh78}
  Bohlin, R. C., Savage, B. D., \& Drake, J. F. 1978, \apj, 224, 132

\bibitem[Bonnet et al.(2004)]{bon04}
  Bonnet et al. 2004, The ESO Messenger, 117, 17

\bibitem[Bradford et al.(2005)]{bra05} 
 Bradford, C. M., Stacey,
  G. J., Nikola, T., Bolatto, A. D., Jackson, J. M., Savage,
  M. L., \& Davidson, J. A.  2005, \apj, 623, 866

\bibitem[Brown \& Liszt(1984)]{bro84}
    Brown, R. L., \& Liszt, H. S. 1984, \araa, 22, 223

\bibitem[Chackerian \& Tipping(1983)]{cha83}
  Chackerian, Jr., C., \& Tipping, R. H. 1983, J. Molec. Spectrosc., 99,
  431

\bibitem[Chiappini(2001)]{chi01} Chiappini, C., Matteucci, F.,
  \& Romano, D. 2001, \apj, 554, 1044

\bibitem[Christopher et al.(2005)]{chr05}
    Christopher, M. H., Scoville, N. Z., Stolovy, S. R., \& Yun,
   M. S. 2005, \apj, 622, 346

\bibitem[Coil \& Ho(2000)]{coi00}
   Coil, A. L., \& Ho, P. T. P. 2000, \apj, 533, 245

\bibitem[Crocker et al.(2011)]{cro11}
 Crocker, R. M., Jones, D. I., Aharonian, F., Law, C. J., Melia, F.,
  Oka, T., \& Ott, J. 2011, MNRAS, 413, 763

\bibitem[Dame \& Thaddeus(2008)]{dam08}
   Dame, T. M. \& Thaddeus, P. 2008, \apjl, 683, L143

\bibitem[Eisenhauer et. al(2003)]{eis03}
  Eisenhauer, F., Sch\"odel, R., Genzel, R., Ott, T., Tecza, M.,
  Abuter, R., Eckart, A., \& Alexander, T. 2003, \apjl, 597, L121

\bibitem[Esteban et al.(2005)]{est05}
  Esteban, C., Garc\'ia-Rojas, J., Peimbert, M., Peimbert, A.,
  Ruiz, M. T., Rodr\'iguez, M., \& Carigi, L.  2005, \apj, 618,
  L95

\bibitem[Geballe \& Oka(1996)]{geb96}
   Geballe, T. R., \& Oka, T. 1996, \nat, 384, 334

\bibitem[Geballe, Baas, \& Wade(1989)]{geb89}
   Geballe, T. R., Baas, F., \& Wade, R. 1989, \aap, 208, 255

\bibitem[Geballe \& Oka(2010)]{geb10} 
Geballe, T. R. \& Oka, T. 2010, \apjl, 709, L70

\bibitem[Geballe et al.(1999)]{geb99}
   Geballe, T. R., McCall, B. J., Hinkle, K. H., \& Oka, T.\ 1999,
   \apj, 510, 251

\bibitem[Genzel \& Townes(1987)]{gen87}
Genzel, R. \& Townes, C. H. 1987, \araa, 25, 377

\bibitem[Gerin et al.(2010)]{ger10}
 Gerin, M., de Luca, M., Black, J., et al. 
2010, \aap, 518, L110

\bibitem[Ghez et al.(2008)]{ghe08}
  Ghez, A. M., Salim, S., Weinberg, N. N., Lu, J. R., Do, T.,
  Dunn, J. K., Matthews, K., Morris, M. R., Yelda, S., Becklin,
  E. E., Kremenek, T., Milosavljevic, M., \& Naiman, J. 2008,
  \apj, 689, 1044

\bibitem[Goicoechea et al.(2013)]{goi13}
 Goicoechea, J. R., Etxaluze, M., Cernicharo, J., Gerin, M.,
Neufeld, D. A., Contursi, A., Bell, T. A., de Luca, M.,
Encrenaz, P., Indriolo, N., Lis, D. C., Polehampton, E. T., \&
Sonnentrucker, P. 2013, \apjl, 769, L13

\bibitem[Goto et al.(2002)]{got02}
    Goto, M., McCall, B. J., Geballe, T. R., Usuda, T., Kobayashi, N.,
    Terada, H., \& Oka, T. 2002, \pasj, 54, 951

\bibitem[Goto et al.(2008)]{got08}
  Goto, M., Usuda, T., Nagata, T., Geballe, T.R., McCall, B. J.,
  Indriolo, N., Suto, H., Henning, Th., Morong, C. P., \& Oka,
  T. 2008, \apj, 688, 306

\bibitem[Goto et al.(2011)]{got11}
Goto, M., Usuda, T., Geballe, T. R., Indriolo, N., McCall, B. J.,
 Henning, Th., \& Oka, T. 2011, \pasj, 63, L13

\bibitem[Goto et al.(2013)]{got13}
Goto, M., Indriolo, N., Geballe, T. R., \& Usuda, T. 2013, 
J. Phys. Chem., 117, 9919

\bibitem[G\"usten, Walmsley, \& Pauls(1981)]{gus81}
G\"usten, R., Walmsley, C. M., \& Pauls, T. 1981, \aap, 103, 197

\bibitem[Ho et al.(1991)]{lho91}
   Ho, L. C., Szczepanski, J. C., Ho, P. T. P., Jackson, J. M., \& Armstrong,
   J. T., 1991, Atoms, ions, and molecules: New Results in spectral line 
   astrophysics, ASP Conference Series (ASP: San Francisco), 16, p 177.


\bibitem[Immer et al.(2012)]{imm12}
Immer, K., Schuller, F., Omont, A., \& Menten, K. M. 
2012, \aap, 537, 121

\bibitem[Indriolo et al.(2007)]{ind07} 
Indriolo, N., Geballe, T. R., Oka, T., \& McCall, B. J. 2007, \apj,
671, 1736

\bibitem[Indriolo \& McCall(2012)]{ind12} 
Indriolo, N., \& McCall, B. J. 2012, \apj, 745, 91


\bibitem[Kaifu et al.(1972)]{kai72}
Kaifu, N., Kato, T., \& Iguchi, T. 1972, Nature Phys. Sci., 238, 105

\bibitem[K\"aufl et al.(2004)]{kau04}
  K\"aufl, H. U., et al. 2004, \procspie, 5492, 1218 

\bibitem[Kramer et al.(2004)]{kra04} 
Kramer, C., Jakob, H., Mookerjea, B., Schneider, N., Br\"ull, M.,
Stutzki, J. 2004, \aap, 424, 887

 \bibitem[Lacy et al.(1994)]{lac94}
     Lacy, J. H., Knacke, R., Geballe, R., \& Tokunaga, A. T. 1994,
\apjl, 428, L69

\bibitem[Lazio \& Cordes(1998)]{laz98}
Lazio, T. J. W., \& Cordes. J. M. 1998, \apj, 505, 715

\bibitem[Lis et al.(2001)]{lis01}
Lis, D. C., Serabyn, E., Zylka, R., \& Li, Y. 2001, \apj,
550, 761

\bibitem[Liszt et al.(1983)]{lis83}
Liszt, H. S., van der Hulst, J. M., Burton, W. B., \& Ondrechen,
M. P. 1983, \aap, 126, 341

\bibitem[Lord(1992)]{lor92}
 Lord, S. D. 1992, A New Software Tool for Computing Earth's
 Atmosphere Transmissions of Near- and Far-Infrared Radiation,
 NASA Technical Memoir 103957 (Moffett Field, CA: NASA Ames
 Research Center)


\bibitem[McCall et al.(2002)]{mcc02} 
    McCall, B. J., Hinkle, K. H., Geballe, T. R., Moriarty-Schieven,
    G. H., Evans, N. J., II, Kawaguchi, K., Takano, S., Smith, V. V.,
    \& Oka, T. 2002, \apj, 567, 391

\bibitem[McCall et al.(2004)]{mcc04}
McCall, B. J. et al. 2004, Phys. Rev. A, 70, 052716

\bibitem[McCall et al.(1999)]{mcc99}
McCall, B. J., Geballe, T. R., Hinkle, K. H., \& Oka, T.
1999, \apj, 522, 338

\bibitem[Mills et al.(2013)]{mil13}
 Mills, E. A. C., G\"usten, R., Requena-Torres, M. A., \& Morris, M. R.
2013, \apj, 779, 47

\bibitem[Menon \& Ciotti(1970)]{men70}
Menon, T. K., \& Ciotti, J. E. 1970, \nat, 227, 579

\bibitem[Moneti, Cernicharo, \& Pardo(2001)]{mon01}
   Moneti, A., Cernicharo, J., \& Pardo, J., R., 2001,
  \apjl, 549, L203

\bibitem[Montero-Casta\~no  et al.(2009)]{mon09}
Montero-Casta\~no, M., Herrnstein, R. M., \& Ho, P. T. P.
2009, \apj, 695, 1477

\bibitem[Morris et al.(1983)]{mor83} 
  Morris, M., Polish, N., Zuckerman, B., \& Kaifu, N. 1983, \aj,
  88, 1228

\bibitem[Morris \& Serabyn(1996)]{mor96}
    Morris, M., \& Serabyn, E. 1996, \araa, 34, 645


\bibitem[Moultaka et al.(2009)]{mou09}
Moultaka, J., Eckart, A., \& Sch\"odel, R. 2009, \apj, 703, 1635

\bibitem[Muno et al.(2009)]{mun09}
Muno, M. P., Bauer, F. E., Baganoff, F. K.,
Bandyopadhyay, R. M., Bower, G. C., Brandt, W. N., Broos, P. S.,
Cotera, A., Eikenberry, S. S., Garmire, G. P., Hyman, S. D.,
Kassim, N. E., Lang, C. C., Lazio, T. J. W., Law, C., Mauerhan,
J. C., Morris, M. R., Nagata, T., Nishiyama, S., Park, S.,
Ram\`irez, S. V., Stolovy, S. R., Wijnands, R., Wang, Q. D.,
Wang, Z., \& Yusef-Zadeh, F. 2009, \apjs, 181, 110

\bibitem[Neale et al.(1996)]{nea96}
Neale, L., Miller, S. \& Tennyson, J. 1996, \apj, 464, 516

\bibitem[Neufeld(2012)]{neu12}
Neufeld, D. A. 2012, \apj, 749, 125

\bibitem[Oka (2006)]{oka06}
    Oka, T. 2006, Proc. Natl. Acad. Sci. USA, 103, 12235

\bibitem[Oka \& Epp(2004)]{oka04}
   Oka, T., \& Epp, E. 2004, \apj, 613, 349

\bibitem[Oka et al.(2005)]{oka05}
   Oka, T., Geballe, T. R., Goto, M., Usuda, T., \& McCall,
   B. J. 2005, \apj, 632, 882

\bibitem[Oka et al.(1998)]{oka98} 
   Oka, T., Hasegawa, T., Sato, F., Tsuboi, M., \& Miyazaki,
   A. 1998, \apjs, 118, 455

\bibitem[Oort et al.(1958)]{oor58}
Oort, J. H., Kerr, F. J., \& Westerhout, G. 1958, \mnras, 118, 379

\bibitem[Pan \& Oka(1986)]{pan86}
 Pan, F.-S., \& Oka, T. 1986, \apj, 305, 518

\bibitem[Paumard et al.(2004)]{pau04} Paumard, T., Genzel, R.,
  Maillard, J.-P., Ott, T., Morris, M. R., Eisenhauer, F., \&
  Abuter, R. 2004, in Young Local Universe, ed. A. Chalabaev et
  al. (Paris: \'Editions Fronti\`eres), 377

 \bibitem[Paumard et al.(2006)]{pau06}
    Paumard, T., Genzel, R., Martins, F., Nayakshin, S.,
    Beloborodov, A. M., Levin, Y., Trippe, S., Eisenhauer, F.,
    Ott, T., Gillessen, S., Abuter, R., Cuadra, J., Alexander, T., \&
    Sternberg, A. 2006, \apj, 643, 1011

\bibitem[Petrignani et al.(2011)]{pet11} Petrignani, A.,
  Altevogt, S., Berg, M. H., Bing, D., Grieser, M., Hoffmann,
  J., Jordon-Thaden, B., Krantz, C., Mendes, M. B., Novotn\'y,
  Old\u{r}ich, Novotny, S., Orlov, D. A., Repnow, R., Sorg, T.,
  St\"utzel, J., Wolf, A., Buhr, H., Kreckel, H., Kokoouline,
  V., \& Greene, C. H. 2011, Phys. Rev. A, 83, 032711

\bibitem[Pierce-Price et al.(2000)]{pie00}
 Pierce-Price, D., Richer, J. S., Greaves, J. S., et al. 2000, \apj,
 545, L121

\bibitem[Requena-Torres et al.(2012)]{req12}
 Requena-Torres, M. A., G\"usten, R, Wei\ss, A., Harris, A. I.,
Mart\'in-Pintado, J., Stutzki, J., Klein, B., Heyminck, S., \&
Risacher, C. 2012, \aap, 542, L21

\bibitem[Rolleston et al.(2000)]{rol00} 
Rolleston, W. R. J., Smartt, S. J., Dufton, P. L., \& Ryans,
R. S. I.  2000, \aap, 363, 537

\bibitem[Sandqvist(1970)]{san70}
Sandqvist, Aa. 1970, \aj, 75, 135


\bibitem[Schilke et al.(2010)]{sch10}
Schilke, P., Comito, C., M\"uller, H. et al. 2010, \aap, 521,
L11

\bibitem[Scoville(1972)]{sco72}
Scoville, N. Z. 1972, \apj, 175, L127

\bibitem[Serabyn \& G\"usten(1986)]{ser86}
Serabyn, E., \& G\"usten, R. 1986, \aap, 161, 334

\bibitem[Sodroski et al.(1995)]{sod95}
Sodroski, T. J., Odegard, N., Dwek, E., Hauser, M. G., Franz,
B. A., Freedman, I., Kelsall, T., Wall, W. F., Berriman, G. B.,
Odenwald, S. F., Bennett, C., Reach, W. T., \& Weiland,
J. L. 1995, \apj, 452, 262

\bibitem[Sofia et al.(2004)]{sof04}
Sofia, U. J., Lauroesch, J. T., Meyer, D. M., \& Cartledge,
S. I. B.  2004, \apj, 605, 272

\bibitem[Sofue(1995)]{sof95}
Sofue, Y. 1995, \pasj, 47, 551

\bibitem[Sutton et al.(1990)]{sut90}
Sutton, E. C., Danchi, W. C., Jaminet, P. A., \& Masson, C. R.
1990, \apj, 348, 503


\bibitem[Van der Tak et al.(2007)]{tak07}
van der Tak, F. F. S., Black, J. H., Sch\"oier, F. L., Jansen,
D. J., van Dishoeck, E. F. 2007, \aap, 468, 627

\bibitem[van Woerden, Rougoor, \& Oort(1957)]{woe57} 
   van Woerden, H., Rougoor, G. W., \& Oort, J. H. 1957, {\it Comptes
   Rendus}, 244, 1691

\bibitem[Wilms, Allen, \& McCray(2000)]{wil02}
Wilms, J., Allen, A., \& McCray, R., 2000, \apj, 542, 914


\bibitem[Yusef-Zadeh et al.(2007)]{yus07a}
 Yusef-Zadeh, F., Muno, M., Wardle, M., \& Lis, D. C. 2007, \apj,
   656, 847

 \bibitem[Yusef-Zadeh, Wardle, \& Roy(2007)]{yus07b} 
  Yusef-Zadeh, F., Wardle, M., \& Roy, S. 2007, \apjl, 665, L123


\bibitem[Zhao et al.(2009)]{zha09}
 Zhao, J., Morris, M. R., Goss, W. M., \& An, T. 2009, \apj, 699, 186

\bibitem[Zou \& Varanasi(2002)]{zou02}
Zou, Q., \& Varanasi, P. 2002, J. Quant. Spectrosc. Rad. Transf. 75,
63

\end{thebibliography}
